\documentclass[12pt]{article}
\usepackage{amsmath, amsthm}
\usepackage{amscd}
\usepackage{amssymb}
\usepackage{array}
\usepackage{color}
\usepackage{youngtab}

\newtheorem{thm}{Theorem}[section]
\newtheorem{theorem}[thm]{Theorem}

\newtheorem{lemma}[thm]{Lemma}
\newtheorem{proposition}[thm]{Proposition}

\theoremstyle{definition}
\newtheorem{definition}[thm]{Definition}
\newtheorem{example}[thm]{Example}

\newtheorem{remark}[thm]{Remark}

\begin{document}

\newcommand{\comment}[1]{{\color{blue}\rule[-0.5ex]{2pt}{2.5ex}}
\marginpar{\scriptsize\begin{flushleft}\color{blue}#1\end{flushleft}}}

\newcommand{\be}{\begin{equation}}
\newcommand{\ee}{\end{equation}}
\newcommand{\ba}{\begin{align}}
\newcommand{\ea}{\end{align}}
\newcommand{\ban}{\begin{align*}}
\newcommand{\ean}{\end{align*}}

\newcommand{\id}{\relax{\rm 1\kern-.28em 1}}
\newcommand{\R}{\mathbb{R}}
\newcommand{\N}{\mathbb{N}}
\newcommand{\C}{\mathbb{C}}
\newcommand{\Z}{\mathbb{Z}}
\newcommand{\g}{\mathfrak{G}}
\newcommand{\e}{\epsilon}

\newcommand{\hs}{\hfill\square}
\newcommand{\hbs}{\hfill\blacksquare}

\newcommand{\bp}{\mathbf{p}}
\newcommand{\bmax}{\mathbf{m}}
\newcommand{\bT}{\mathbf{T}}
\newcommand{\bU}{\mathbf{U}}
\newcommand{\bP}{\mathbf{P}}
\newcommand{\bA}{\mathbf{A}}
\newcommand{\bm}{\mathbf{m}}
\newcommand{\bIP}{\mathbf{I_P}}

\newcommand{\cA}{\mathcal{A}}
\newcommand{\cB}{\mathcal{B}}
\newcommand{\cC}{\mathcal{C}}
\newcommand{\cI}{\mathcal{I}}
\newcommand{\cO}{\mathcal{O}}
\newcommand{\cG}{\mathcal{G}}
\newcommand{\cJ}{\mathcal{J}}
\newcommand{\cF}{\mathcal{F}}
\newcommand{\cP}{\mathcal{P}}
\newcommand{\ep}{\mathcal{E}}
\newcommand{\E}{\mathcal{E}}
\newcommand{\cH}{\mathcal{O}}
\newcommand{\cPO}{\mathcal{PO}}
\newcommand{\cl}{\ell}
\newcommand{\cFG}{\mathcal{F}_{\mathrm{G}}}
\newcommand{\cHG}{\mathcal{H}_{\mathrm{G}}}
\newcommand{\Gal}{G_{\mathrm{al}}}
\newcommand{\cQ}{G_{\mathcal{Q}}}
\newcommand{\cT}{\mathcal{T}}
\newcommand{\cM}{\mathcal{M}}

\newcommand{\ri}{\mathrm{i}}
\newcommand{\re}{\mathrm{e}}
\newcommand{\rd}{\mathrm{d}}
\newcommand{\rSt}{\mathrm{St}}
\newcommand{\rGL}{\mathrm{GL}}
\newcommand{\rSU}{\mathrm{SU}}
\newcommand{\rSL}{\mathrm{SL}}
\newcommand{\rSO}{\mathrm{SO}}
\newcommand{\rOSp}{\mathrm{OSp}}
\newcommand{\rSpin}{\mathrm{Spin}}
\newcommand{\rsl}{\mathrm{sl}}
\newcommand{\rM}{\mathrm{M}}
\newcommand{\rU}{\mathrm{U}}
\newcommand{\rdiag}{\mathrm{diag}}
\newcommand{\rP}{\mathrm{P}}
\newcommand{\rdeg}{\mathrm{deg}}
\newcommand{\rStab}{\mathrm{Stab}}
\newcommand{\rcof}{\mathrm{cof}}

\newcommand{\M}{\mathrm{M}}
\newcommand{\End}{\mathrm{End}}
\newcommand{\Hom}{\mathrm{Hom}}
\newcommand{\diag}{\mathrm{diag}}
\newcommand{\rspan}{\mathrm{span}}
\newcommand{\rank}{\mathrm{rank}}
\newcommand{\Gr}{\mathrm{Gr}}
\newcommand{\ber}{\mathrm{Ber}}

\newcommand{\fsl}{\mathfrak{sl}}
\newcommand{\fg}{\mathfrak{g}}
\newcommand{\ff}{\mathfrak{f}}
\newcommand{\fgl}{\mathfrak{gl}}
\newcommand{\fosp}{\mathfrak{osp}}
\newcommand{\fm}{\mathfrak{m}}

\newcommand{\ttau}{\tilde\tau}

\newcommand{\str}{\mathrm{str}}
\newcommand{\Sym}{\mathrm{Sym}}
\newcommand{\tr}{\mathrm{tr}}
\newcommand{\defi}{\mathrm{def}}
\newcommand{\Ber}{\mathrm{Ber}}
\newcommand{\spec}{\mathrm{Spec}}
\newcommand{\sschemes}{\mathrm{(sschemes)}}
\newcommand{\sschemeaff}{\mathrm{ {( {sschemes}_{\mathrm{aff}} )} }}
\newcommand{\rings}{\mathrm{(rings)}}
\newcommand{\Top}{\mathrm{Top}}
\newcommand{\sarf}{ \mathrm{ {( {salg}_{rf} )} }}
\newcommand{\arf}{\mathrm{ {( {alg}_{rf} )} }}
\newcommand{\odd}{\mathrm{odd}}
\newcommand{\alg}{\mathrm{(alg)}}
\newcommand{\sa}{\mathrm{(salg)}}
\newcommand{\sets}{\mathrm{(sets)}}
\newcommand{\SA}{\mathrm{(salg)}}
\newcommand{\salg}{\mathrm{(salg)}}
\newcommand{\varaff}{ \mathrm{ {( {var}_{\mathrm{aff}} )} } }
\newcommand{\svaraff}{\mathrm{ {( {svar}_{\mathrm{aff}} )}  }}
\newcommand{\ad}{\mathrm{ad}}
\newcommand{\Ad}{\mathrm{Ad}}
\newcommand{\pol}{\mathrm{Pol}}
\newcommand{\Lie}{\mathrm{Lie}}
\newcommand{\Proj}{\mathrm{Proj}}
\newcommand{\rGr}{\mathrm{Gr}}
\newcommand{\rFl}{\mathrm{Fl}}
\newcommand{\rPol}{\mathrm{Pol}}
\newcommand{\rdef}{\mathrm{def}}
\newcommand{\rE}{\mathrm{E}}

\newcommand{\uspec}{\underline{\mathrm{Spec}}}
\newcommand{\uproj}{\mathrm{\underline{Proj}}}

\newcommand{\sym}{\cong}

\newcommand{\al}{\alpha}

\newcommand{\lam}{\lambda}
\newcommand{\de}{\delta}
\newcommand{\D}{\Delta}
\newcommand{\s}{\sigma}
\newcommand{\lra}{\longrightarrow}
\newcommand{\ga}{\gamma}
\newcommand{\ra}{\rightarrow}

\newcommand{\tit}{t}
\newcommand{\ts}{\tilde{s}}
\newcommand{\ty}{\tilde{y}}
\newcommand{\titp}{{t}^{\, '}}
\newcommand{\hv}{ \hat{t}}
\newcommand{\hy}{\hat{y}}
\newcommand{\hS}{\hat{S}(x)}
\newcommand{\hx}{\hat{x}}
\newcommand{\hT}{\hat{T}}

\newcommand{\NOTE}{\bigskip\hrule\medskip}


\smallskip

    \centerline{\LARGE \bf  Quantum twistors }

\vskip 2cm

\centerline{ D. Cervantes$^1$, R. Fioresi$^2$,  \underline {M. A. Lled\'{o}}$^{3,4}$,  and Felip A. Nadal$^4$}

\vskip 2cm

 \centerline{\it $^1$  Computer Science Department, CINVESTAV-IPN,}
\centerline{ \it
Mexico City, Mexico.}

\bigskip

\centerline{\it $^2$ Dipartimento di Matematica,  Universit\`{a} di
Bologna,}
\centerline{ \it Piazza di Porta S. Donato, 5. 40126 Bologna. Italy.}

\bigskip

 \centerline{\it $^3$  Departament de F\'{\i}sica Te\`{o}rica,
Universitat de Val\`{e}ncia.}
\centerline{\it $^4$  Institut de F\'{\i}sica Corpuscular  (CSIC-UVEG).}
\centerline{\small\it C/Dr.
Moliner, 50, E-46100 Burjassot (Val\`{e}ncia), Spain.}

\vskip 1cm

 \centerline{{\footnotesize e-mail: dalia@computacion.cs.cinvestav.mx, fioresi@dm.unibo.it,}} \centerline{{\footnotesize Maria.Lledo@ific.uv.es,  Felip.Nadal@gmail.com}}

\vskip 2cm

\begin{abstract}
We compute explicitly a star product on the Minkowski space whose Poisson bracket is quadratic. This star product corresponds to a deformation  of the conformal spacetime, whose big cell is the Minkowski spacetime. The description of Minkowski space is made in the twistor formalism and the quantization follows by substituting the classical conformal group by a quantum group.
\end{abstract}

\vfill\eject

\section{Introduction}

Twistor geometry \cite{pe,pm,ww} arose as an alternative way of describing spacetime. One starts with an abstract four dimensional complex vector space ({\it twistor space}) and the complex, compactified Minkowski space is seen as the set of two planes inside the twistor space. This is the Grassmannian manifold $G(2,4)$, and it is a homogeneous space of the group $\rSL(4,\C)$, which is  the complexification of $\rSU(2,2)$,  the spin  (two fold) cover of the conformal group $\rSO(2, 4)$. So we can properly call $G(2,4)$ the {\it conformal space}.

There is an obvious advantage of the formulation: the action of the conformal group is explicit, since it comes naturally into play right at the beginning of the construction. Conformal invariance is not a symmetry of all the physical theories (it is a symmetry of electromagnetism, for example), so it should be an explicitly broken symmetry. As pointed out in Ref. \cite{pm}, one can write down any field theory  in the twistor formalism and then the terms that break the invariance appear isolated. This could then clarify the mechanisms for the explicit breaking of the symmetry.  In mathematical terms, one passes from the Minkowski space to conformal space by a {\it compactification} and viceversa by restricting to the {\it big cell} of the conformal space. So one could think on a non conformally symmetric field theory as a conformal theory broken down to the big cell by some extra terms.

Moreover, conformal symmetry has a fundamental fundamental role in the gauge/gravity correspondence \cite{mal1} (for a  review see Refs.  \cite{mal2,ae} ) which relates gravity theories to conformally invariant gauge theories defined on a boundary of spacetime.

In the original papers \cite{pe,pm}, Penrose believed that twistor theory could help to introduce the indetermination principle in spacetime. The  points had to be `smeared out' and in twistor formalism a point of spacetime is not a fundamental quantity, but it is secondary to twistors.

Nevertheless, all the twistor construction is classical. Our point of view is introducing   the quantum indetermination principle in spacetime by deforming the algebra of functions over spacetime to a noncommutative algebra. An example of deformation are the {\it quantum groups} \cite{frt}, a non commutative deformation of algebraic Lie groups. In algebraic terms, a group is retrieved through its function algebra, which is a commutative but  non cocommutative {\it Hopf algebra}.  Quantum groups are non commutative, non cocommutative  Hopf algebras depending on an indeterminate parameter $q$. One can specify $q=1$ to recover the original commutative Hopf algebra, or to any real or complex value to obtain examples of non commutative Hopf algebras.

The quantum group   $\rSL_q(4,\C)$ would then be the the complexified  quantum conformal group. The idea underlying the work of Refs. \cite{fi2,fi3} was to make such  substitution and then to obtain a quantum Grassmannian, a quantum Minkowski space and a quantum Poincar\'{e} group satisfying the same relations among them as their classical counterparts. So the quantum conformal group acts naturally on the quantum Grassmannian, viewed as a quotient, and  the quantum Poincar\'{e} group is identified with the subgroup of it that preserves the big cell.  This construction has also been generalized to flag manifolds \cite{fi4}.

In the super setting, we have several superspaces that are of interest: the Grassmannian supervariety $\rGr(2|0,4|1)$, which corresponds in physical terms to the algebra of {\it chiral superfields}  and the superflag $Fl(2|0, 2|1,4|1)$ which is the complexification of the $N=1$ {\it Minkowski superspace}. The same idea can be applied here with the supergroup $\rSL(4|1)$ \cite{cfl1,cfl2}, which also can be deformed to a quantum supergroup. For a detailed treatment of all, the super and  non super, classical and quantum  cases see Ref. \cite{fl3}.

Here we deal only with the non super, quantum case. We have identified a quantization of the conformal space  as an homogeneous space of $\rSL_q(4,\C)$. This quantization can be given in  more concrete terms. In the big cell (the Minkowski space) it can be presented as a {\it star product} on the algebra of functions. There is an atlas of the Grassmannian with 6 identical cells, and the star products in the intersections glue in such way that one can recover the quantum Grassmannian.

We are working in the algebraic category, so we first give an explicit formula for the star product   among two polynomials in the big cell of the Grassmannian. Since the quantum algebras that we present here are deformations of the algebra of polynomials on Minkowski space, the star product that we obtain is algebraic.

We then show  that this deformation can be extended  to the set of smooth functions in terms of a differential star product. The Poisson bracket (the antisymmetrized first order term  in $h$ with $q=\re^h$) of the deformation is a quadratic one, so the Poisson structure is not symplectic (nor regular).

 Examples of such  transition  from the category of algebraic varieties to the category of differential manifolds in the quantum theory are given in Refs. \cite{fl1,ll,fll,fl2}. In these references, the varieties under consideration are coadjoint orbits and the Poisson bracket is linear. It was shown in that paper that some algebraic star products do not have differential counterpart (not even modulo and equivalence transformation), so the results of this paper are non trivial.  It is interesting that one of the algebraic star products that does not have differential extension   is the star product on the coadjoint orbits of $\rSU(2)$, associated to the standard quantization of angular momentum. For algebraic star products and their classification, see also Ref. \cite{ko2}.

 \medskip

There are previous works that deal with the quantization of space time in terms of the twistor space. One has, for example, the interesting relation of twistors with geometric quantization in Ref. \cite{ch}. More recently, in  Ref. \cite{ha}, the authors introduce first a constant, symplectic form on the Minkowski space which gives rise to a Weyl-Moyal deformation. A deformation of the conformal group through the R-matrix approach is considered in order to construct the action of the conformal group on the noncommutative space. As the authors claim, the resulting deformation is the same than the one used in \cite{kko}. The Moyal deformation of space time has been used in string theory (the original references are Refs. \cite{cds,sw}). The origin of the symplectic form is a $B$-field (an antisymmetric, 2-tensor field) that acquires in some backgrounds a constant vacuum expectation value. This constant, antisymmetric matrix can be interpreted as a Poisson structure on the Minkowski space and the Weyl-Moyal quantization or star product is then a genuine noncommutative structure for spacetime. The Weyl-Moyal star product is, in some sense, the simplest formal deformation that one can construct on $\R^n$. It requires a constant Poisson bracket:
$$\{f(x), g(x)\}= B^{\mu\nu}\partial_\mu f (x)\partial_\nu g (x),\qquad f, g\in  C^\infty(\R^n)\,,$$ where $B^{\mu\nu}$ is a constant antisymmetric matrix
and then the Weyl-Moyal formal deformation  or star product is
$$f\star g (x)=\sum_{n=0}^\infty \frac{h^n}{n!}B^{\mu_1\nu_1}\cdots
B^{\mu_n\nu_n}\partial_{\mu_1}\dots\partial_{\mu_n}f(x)
\partial_{\nu_1}\dots\partial_{\nu_n}g(x)\,.$$
The symbol $h$ is the formal parameter of the deformation.
There are very few deformations that can be given explicitly in closed form. A general formula is known for an arbitrary Poisson bracket (Kontsevich's formula, \cite{ko}) but it is extremely hard to work out the coefficients for the differential operators appearing in the deformation, even for simple, linear Poisson brackets.

Another approach is to take advantage of the fact that the Grassmannian $G(m, n)\simeq \rSL(n)/S\left(\rGL(m)\times \rGL(n-m)\right)$ is a coadjoint orbit of the group $SL(n)$. In fact, any flag manifold is so, being the  full flag $Fl(1,2,3,\dots,n)$ the regular (maximal dimension) orbit and all the others non regular. The approach of Refs. \cite{fl1,ll}  would then be relevant here. The {\sl Kirillov-Kostant-Soriau Poisson bracket} on the coadjoint algebra restricts to a symplectic Poisson bracket on the orbits. It is essentially given by the Lie bracket and the star product is obtained from the enveloping algebra. It is then an {\sl equivariant} star product under the action of the group. In these works the quantization is given in terms of generators and relations so it is  algebraic, but then in Refs. \cite{fll,fl2} the relation with differential star products was studied.

Another approach to the quantization of coadjoint orbits has been undertaken also in Refs. \cite{al,eem,mu} using the so-called {\it Shapovalov pairing } of Verma modules.

Grassmannians have also be quantized as {\sl fuzzy spaces}. This means that one uses harmonic functions on the coset space and the expansion is truncated at some level. The functions can then be seen as matrices and a product on the truncated space is defined just using matrix multiplication. We find this approach in Refs. \cite{dm,dj}.

We believe that the three approaches just mentioned must be linked in some way, since the quantizations are equivariant under the classical group ($\rSL(4,\C)$ in this case) and all of them are intimately related to representation theory. It is, however, not straightforward to compare them.

Interesting as these works are, our deformation is a different one. The Poisson bracket that we obtain on the Minkowski space is a quadratic one (in particular, not symplectic) and the star product is then non equivalent to a Weyl-Moyal one. Also, the equivariance of the star product is achieved only by deforming the group to a quantum group, contrary to the above mentioned approaches. Nevertheless, we are able to give an explicit formula for it in terms of a recursive expression. This deformation could eventually have a similar interpretation in string theory considering a non constant background field $B$. We have not explored yet that possibility.

\medskip

 The organization of the paper is as follows:

 In Section \ref{sec:Grassmannian} we review the classical picture and settle the notation for the algebraic approach. In Section \ref{sec:The quantum} we describe the quantum Minkowski space obtained in Refs. \cite{fi2,fi3,cfl1, cfl2}, together with the corresponding quantum groups. In Section \ref{sec:Algebraic} we give the explicit formula for the star product among two polynomials on Minkowski space. In Section \ref{sec:Differential} we prove that the star product can be extended to smooth functions and compute it explicitly up to order two in $h$. In Section \ref{sec:Poincare} we show that the coaction of the Poincar\'{e} group on the quantum Minkowski space is representable by a differential operator (at least up to order one in $h$). To show this, we need a  technical result concerning the quantum Poincar\'{e} group, that   we prove  in the appendix \ref{diamond}.
 In Section \ref{sec:The real} we study the real forms of the quantum algebras that correspond to the real forms of ordinary Minkowski and Euclidean space. In Section \ref{sec:The deformed} we write the quadratic invariant (the metric of the Minkowski space) in the star product algebra. Finally, in Section \ref{sec:Conclusions} we state our conclusions and outlook.

\section{Grassmannian, conformal group and \\Minkowski space.} \label{sec:Grassmannian}

We give here the classical description of the conformal space as a Grassmannian variety and the Minkowski space as the big cell inside it. This description is well known (see for example Refs. \cite{ww,va}). We follow closely the notation of Refs. \cite{va,flv,cfl1,cfl2,fl3}.

\begin{definition}The {\it conformal space} is the  Grassmannian variety $G(2,4)$, the set of 2-planes inside a four dimensional space $\cT\simeq\C^4$, which is called the {\it twistor space}. \hfill $\blacksquare$\end{definition}

 A plane $\pi$ in $\cT$ can be  given by two linearly
independent vectors
$$\pi=(a,b)=\rspan\{a,b\},\qquad a, b \in\cT.$$
If $\rspan\{a,b\}=\rspan\{a',b'\}$ they define the same
point of the Grassmannian.  This means that we can take linear combinations of the vectors $a$ and $b$
\be(a',b')=(a,b)h,\qquad \qquad h\in \rGL(2, \C)\label{GL2}\ee
 to represent the same plane $\pi$.
 In the following, we will use the identification of $\cT$ with $\C^4$. Then,  in the canonical basis, we have
$$ \begin{pmatrix}a'_1&b'_1\\a'_2&b'_2\\a'_3&b'_3\\a'_4&b'_4\\\end{pmatrix}=\begin{pmatrix}a_1&b_1\\a_2&b_2\\a_3&b_3\\a_4&b_4\\\end{pmatrix} \begin{pmatrix}h_{11}&h_{12}\\h_{21}&h_{22}\end{pmatrix}\,.$$

What relates the Grassmannian to the conformal group is that there is  a transitive action of
$\rGL(4,\C)$ on $G(2,4)$
$$g\in \rGL(4,\C),\qquad g\pi=(ga,gb).$$ One can take $\rSL(2,\C)$ instead and the action is still transitive. Then, the Grassmannian is a homogeneous space of $\rSL(4,\C)$. Let us take the plane
$$\pi_0=(e_1, e_2)=\begin{pmatrix}1&0\\0&1\\0&0\\0&0\end{pmatrix}\,.$$
 The stability group of $\pi_0$  is the upper parabolic subgroup
\be P_0=\left\{\begin{pmatrix}L&M \\
0&R\end{pmatrix}\in \rSL(4,\C)\quad \big|\quad \det L \cdot \det R=1 \right\}\,,\label{ups}\ee with
$L, M, R$ being $2\times 2$ matrices. Then one has the following result:

\begin{proposition} $G(2,4)$ is the homogeneous space
$$G(2,4)=\rSL(4,\C)/P_0\, ,$$  with $P_0$  the upper parabolic subgroup (\ref{ups}).
\hfill $\blacksquare$ \end{proposition}

The conformal group in  dimension four and Minkowskian signature is the orthogonal group $\rSO(2,4)$. Its spin group is $\rSU(2,2)$. If we consider the complexification, $\rSO(6,\C)$ (later on we will study  the real forms), the spin group is $\rSL(4,\C)$. We have then that the spin group of the complexified conformal group acts transitively on the Grassmannian $G(2,4)$.

We consider now the standard open covering of  $G(2,4)$. As we have seen, a plane $\pi=(a,b)$ can be represented by a matrix
 $$\pi=\begin{pmatrix}a^1&b^1\\a^2&b^2\\
a^3&b^3\\a^4&b^4\end{pmatrix}\,.$$ This matrix has rank two, since the two vectors are independent. So at least one of the $2\times 2$ blocks has to have determinant different from zero. We define the six open sets
\be U_{AB}=\left\{(a,b)\in \C^4\times \C^4\;\big |\; a^Ab^B-b^Ba^A\neq
0\right\},\quad i<j, \quad A, B=1,\dots 4. \label{opensets}\ee
 This is an open covering of $G(2,4)$ by dense open sets. BVy convention, one chooses the set $U_{12}$, which is called the  {\it big cell}  of $G(2,4)$. By using the freedom (\ref{GL2}) we can always bring a plane in $U_{12}$ to the form
\be \pi=\begin{pmatrix}1&0\\0&1\\t_{31}&t_{32}\\t_{41}&t_{42}\end{pmatrix},\label{minkowskicoord}\ee with the entries  of $t$ totally arbitrary. So $U_{12}\approx \C^4$. This leads us to the following definition:

 \begin{definition} The {\it complexified Minkowski space} is the  big cell inside the Grassmannian. We denote it as $\M:=U_{12}$. \hfill $\blacksquare$\end{definition}

 The Poincar\'{e} group is a subgroup of the conformal group and it should act on the Minkowski space. It is not hard to prove the following proposition:

\begin{proposition} The subgroup of $ \rSL(4,\C)$ that leaves invariant the big cell $U_{12}$ consists
of all the matrices of the form
$$P=\left\{\begin{pmatrix}x&0\\Tx&y\end{pmatrix}\quad \big| \quad  \det x\cdot\det y=1\right\}.$$ This is the {\it Poincar\'{e} group times dilations}. \end{proposition}

{\sl Proof.} First of all, notice that the bottom
left entry is arbitrary but we have written it in this way for convenience. The action on $U_{12}$ is then
\begin{equation}\begin{CD}t@>>> ytx^{-1}+T\,,\end{CD}\label{poincare}\end{equation} so $P$ has the structure of semidirect
product $P=H\ltimes M_2$, where $M_2=\{T\}$ is the set of $2\times 2$ matrices,  acting as translations, and
$$H=\left\{\begin{pmatrix}x&0\\0&y\end{pmatrix},\; \; x,y\in
\rGL(2,\C),\;\; \det\!x\cdot \det\!y=1\right\}.$$ The subgroup $H$
is the direct product  $\rSL(2,\C)\times \rSL(2,\C)\times
\C^\times$. But $\rSL(2,\C)\times \rSL(2,\C)$ is the spin group of
$\rSO(4,\C)$, the complexified Lorentz group, and $\C^\times$ acts
as a dilation. We then conclude that $P$ is then the Poincar\'{e} group times dilations.\hfill$\blacksquare$

\smallskip

In the basis of the Pauli matrices
\be
\sigma_{0}=\begin{pmatrix}1&0\\0&1\end{pmatrix},\quad
\sigma_{1}=\begin{pmatrix}0&1\\1&0\end{pmatrix},\quad
\sigma_{2}=\begin{pmatrix}0&-\ri\\\ri&0\end{pmatrix},\quad
\sigma_{3}=\begin{pmatrix}1&0\\0&-1\end{pmatrix},\label{Pauli}
\ee
an arbitrary matrix $t$ can be written as
$$
t=\begin{pmatrix}t_{31}&t_{32}\\t_{41}&t_{42}\end{pmatrix}=x^{0}\sigma_{0}+x^{1}\sigma_{1}+x^{2}\sigma_{2}+x^{3}\sigma_{3}=
\begin{pmatrix}
x^{0}+x^{3} & x^{1}-\ri x^{2} \\
x^{1}+ix^{2} & x^{0}-x^{3}
\end{pmatrix}.
$$
Then $$\det t=(x^0)^2-(x^1)^2-(x^2)^2-(x^3)^2\,.$$
  The quadratic form $\det t$ is left invariant  by the action of the subgroup $\rSL(2,\C)\times \rSL(2,C) \subset P$, which is the spin group od the complexified Lorentz group $\rSO(4,\C)$. This is enough to interpret $(x^0,x^1,x^2,x^3)$  as the ordinary coordinates of the Minkowski space.

Notice that although both, the twistor space $\cT$ and the Minkowski space $\M$ are isomorphic to $\C^4$, they are different spaces and play different roles in the construction.

 We say that the Grassmannian $G(2,4)$ is the  {\it conformal compactification} of the complex Minkowski space. This compactification consists of adding to the Minkowski space a variety of points at infinity. In fact, the set of points that we add are the closure of a cone in $\C^4$ \cite{flv}.

\paragraph       {Algebraic approach.} In the quantum theory the word quantization means changing (or  deforming) the algebra of observables (usually functions over a phase space) to a non commutative one (usually operators over a Hilbert space). Also here, when talking about quantum spacetime we refer to a deformation of a commutative algebra to a non commutative one. The algebra of departure is the algebra of functions over spacetime. We will consider first polynomials (all the objects described above are algebraic varieties). In Section \ref{sec:Differential} we will see how the construction can be extended to smooth functions.

We fist consider the group $\rGL(4,\C)$, an algebraic group. An element of it is, generically,
$$g=\begin{pmatrix}
 g_{11} &  g_{12} &g_{13} &  g_{14}  \\
 g_{21} &  g_{22} & g_{23} &  g_{24} \\
 g_{31} &  g_{32} &  g_{33} &  g_{34}  \\
 g_{41} &  g_{42} &  g_{43} &  g_{44}  \\
\end{pmatrix},\qquad \det g\neq 0\,.$$

The algebra of polynomials of $\rGL(4,\C)$ is the algebra of polynomials in the entries of the matrix, and an extra variable $d$, which then is set to be the inverse of  the determinant, thus forcing the determinant to be different from zero:
$$\cO(\rGL(4,\C))=\C[g_{AB}, d]/ (d\cdot \det g-1),\qquad A,B=1,\dots , 4.$$ If we want to consider the algebra of
$\rSL(4,\C)$ we  have simply
\be \cO(\rSL(4,\C))=\C[g_{AB}]/ (\det g-1),\qquad A,B=1,\dots , 4.\label{sl4}\ee
In both cases the group law is expressed algebraically as a coproduct, given on the generators as
\be\begin{CD}\cO(\rGL(4,\C))@>\Delta>> \cO(\rGL(4,\C))\otimes \cO(\rGL(4,\C))\\g_{AB}@>>>\sum_Cg_{AC}\otimes g_{CB},\\ d@>>>d\otimes d\end{CD}\qquad A,B, C=1,\dots,4,\label{coproduct}\ee and extended by multiplication to the whole $\cO(\rGL(4,\C))$. The coproduct is {non cocommutative}, since switching the two factors of $\Delta f$ does not leave the result unchanged.

We also have the antipode $S$, (which corresponds to the inverse in $\rGL(4,\C)$),
\be\begin{CD}\cO(\rGL(4,\C))@>S>>\cO(\rGL(4,\C))\\g_{AB}@>>>{g^{-1}}_{AB}=d\,(-1)^{B-A}M_{BA}\\d@>>> \det{g}\,,\end{CD}\label{antipode}\ee where $M_{BA}$ is the minor of the matrix $g$ with the row $B$ and the column $A$ deleted. There is compatibility of these maps. For example, one has
\be\Delta (f_1f_2)=\Delta f_1\Delta f_2,  \label{compatibility}\ee
 as well as the properties of {\it associativity} and {\it coassociativity} of the product and the coproduct. There is also a {\it unit} and a {\it counit} (see Ref. \cite{ka}, for example), and all this gives to  $\cO(\rGL(4,\C))$ the structure of a {\it commutative, non cocommutative Hopf algebra.}

\begin{remark}\label{interpretation}

 Let us see intuitively why the coproduct corresponds to the matrix multiplication on the group itself. We try now to see an element of $\cO(\rGL(4,\C))$ as a function over the variety of the group itself.   Let us denote the natural injection
$$\begin{CD}\cO\big(\rGL(4,\C)\big)\otimes \cO\big(\rGL(4,\C)\big)@>  \mu_G>>\cO\big(\rGL(4,\C)\times \rGL(4,\C)\big)\\f_1\otimes f_2@>>>f_1\times f_2\end{CD}$$
such that $f_1\times f_2(g_1,g_2)=f_1(g_1)f_2(g_2).$
Then we have that
$$\mu_G\circ(\Delta f) (g_1, g_2)=f(g_1g_2),\qquad f\in \cO(\rGL(4,\C)).$$ \hfill $\blacksquare$\end{remark}
\begin{remark}\label{remarkcoactionGL}
Since $\cT$ is isomorphic to the affine space $\C^4$, we have that the polynomial algebra on $\cT$ is
$$\cO(\cT)\simeq \C[a^1, a^2, a^3, a^4]\,. $$ The left action\footnote{One can define a right action by multiplying a row vector on the right by the group matrix.} of $\rGL(4,\C)$ on $\cT$ is the fundamental representation
$$\begin{CD}\rGL(4,\C)\times \cT @>>>\cT\\
(g, a)@>>>ga\,.\end{CD}$$
It is expressed in the canonical basis  $\{e_A,\; A=1,\dots ,4\}$ as
$$(ga)_Be_B=g(e_A)_Ba^A= g_{BA}a^Ae_B,\qquad\hbox{where}\quad  a=a^Ae_A\,.$$
In the language of algebras this is translated to a coaction of $\cO\big(\rGL(4,\C)\big)$ on $\cO(\cT)$. Seeing the coordinates $a^A$ as polynomial functions on $\cT$, the left\footnote{One can also define a right coaction.} coaction $\tilde \Delta$ is
\be\begin{CD} \cO(\cT)@>\tilde \Delta>>\cO\big(\rGL(4,\C)\big)\otimes\cO(\cT)\\
 a^A@>>> \sum_Bg_{AB}\otimes a^B\,.\end{CD}\label{coactionGL}\ee
  and, as in Remark \ref{interpretation}, if
\be\begin{CD}\cO\big(\rGL(4,\C)\big)\otimes\cO(\cT)@>\mu_{G\times \cT}>>\cO\big(\rGL(4,\C)\times \cT\big)\end{CD}\label{mu}\ee is the natural injection, then
$$\mu_{G\times \cT}\circ\tilde\Delta(f)(g,  a)=f( g a)\,. $$

\hfill$\blacksquare$
\end{remark}

We deal  with the subgroups $\rSL(4,\C)$ and  $P$ in the same way, being the coproduct and the antipode well defined on their algebras, that is,  on (\ref{sl4}) and
\begin{align}&\cO(P)=\C[x_{ij},  y_{a b},  T_{ai}]/( \det x\cdot \det y -1), \label{opl}\qquad i,j=1,2,\quad  a,b=3,4.\end{align}
Since we have made a change of generators in $P$, we want to express the coproduct and the antipode in terms of $x$, $y$ and $T$:

\begin{align}
&\Delta x_{ij}=x_{ik}\otimes x_{kj},\nonumber\\
&\Delta y_{ab}=y_{ac}\otimes y_{cb},\nonumber\\
&\Delta T_{ai}=T_{ai}\otimes 1+y_{ac}S(x_{ji})\otimes T_{cj}.\label{coproductPl}
\end{align}

\begin{align}
&S(x_{ij})={x^{-1}}_{ij}= \det y\, (-1)^{j-i}M_{ij},\nonumber\\
&S(y_{ij})={y^{-1}}_{ij}=\det x\, (-1)^{j-i}M_{ij},\nonumber\\
&S(T_{ai})=-S(y_{ab})T_{bj} x_{ji}.
\label{antipodePl}
\end{align}

The Minkowski space is isomorphic to the affine space $\C^4$, so its algebra of polynomials is
$$\cO(\M)\simeq\C[ t_{ai}],\qquad a=3,4, \quad i=1,2.$$ The action of the Poincar\'{e} group on the Minkowski space is expressed as a coaction on its algebra
\be\begin{CD} \cO(\M)@>\tilde \Delta>>\cO(P)\otimes\cO(\M)\\
 t_{ai}@>>> y_{ab}S( x)_{ji}\otimes t_{bj}+ T_{ai}\otimes 1.\end{CD}\label{coactionPl}\ee
This corresponds to the standard action (\ref{poincare}).

\hfill $\blacksquare$

\section{The quantum Minkowski space} \label{sec:The quantum}

The quantization of Minkowski and conformal spaces starts with the quantization of $\rSL(4,\C)$. We substitute the group by the corresponding quantum group $\rSL_q(4,\C)$, which is the quantization of the algebra $\cO(\rSL(4,\C))$ and then we quantize the rest of the structures in order to preserve the relations among them. This approach is followed in the series of papers \cite{fi2,fi3,fi4} and we are not reproducing it here. We will only state the result for the quantization of the algebra of Minkowski space. For the proofs, we refer to those papers or to Ref. \cite{fl3}. It is nevertheless important to remind the structure of the quantum group $\rSL_q(4,\C)$.

\begin{remark} If $k$ is a field, we denote by $k_q$ the ring of formal power series in the indeterminates $q$ and $q^{-1}$, with $qq^{-1}=1$.
\hfill $\blacksquare$\end{remark}

 \begin{definition} The {\it quantum twistor space} is the algebra over $\C$ in four indeterminates $\hat a^A$, $A=1,\dots, 4$ with commutation relations
\be\hat a^A\hat a^B- q^{-1}\hat a^B\hat a^A=0,\qquad A<B\,,\label{quantumtwistorrel}\ee that is, the algebra
$$\C^4_q:=\C_q\langle\hat a^1,\dots ,\hat a^4\rangle/(\hat a^A\hat a^B- q^{-1}\hat a^B\hat a^A),\qquad A<B, \qquad A,B=1,\dots,4\,,$$
where $\C_q\langle\hat a^1,\dots ,\hat a^4\rangle$ is the free algebra over the ring $\C_q$ generated by the four variables $\hat a^1,\dots ,\hat a^4$.\hfill$\blacksquare$\end{definition}
This is the four dimensional {\it quantum space} as defined by Manin \cite{ma}.
For $q=1$ we just obtain the algebra of polynomials on $\C^4$. So $\C^4_q$ is a deformation of such polynomial algebra, $\cO(\cT)$. This is why we call it the {\it quantum twistor space}. We can denote it also as $\cO_q(\cT)$

One wants now to define left and right coactions on the quantum twistor space in a way that for $q=1$ the coaction becomes (\ref{coactionGL}).
In order to do that, we first need the following definition.

\begin{definition} A {\it quantum matrix} is a square matrix of indeterminates
$$\hat g=\begin{pmatrix}
 \hat g_{11} &  \hat g_{12} &\hat g_{13} &  \hat g_{14}  \\
 \hat g_{21} &  \hat g_{22} & \hat g_{23} &  \hat g_{24} \\
 \hat g_{31} &  \hat g_{32} &  \hat g_{33} &  \hat g_{34}  \\
 \hat g_{41} &  \hat g_{42} &  \hat g_{43} &  \hat g_{44}  \\
\end{pmatrix}\,,$$
satisfying the Manin relations \cite{ma}
\begin{align}&\hat g_{AB}\;\hat g_{CB}=q^{-1}\hat g_{CB}\;\hat g_{AB}\quad \hbox{ if } A<C,
\nonumber\\
& \hat g_{AB}\; \hat g_{AD}=q^{-1}\hat g_{AD}\;\hat g_{AB},\quad \hbox{ if } B<D,\nonumber\\&\hat g_{AB}\;\hat g_{CD}=\hat g_{CD}\;\hat g_{AB}\quad \hbox{ if } A<C \hbox{ and } D<B \; \hbox{ or } A>C \hbox{ and } D>B,\nonumber\\
&\hat g_{AB}\;\hat g_{CD}-\hat g_{CD}\;\hat g_{AB}=(q^{-1}-q)\;\hat g_{AC}\;\hat g_{BD}\quad \hbox{ if } A<C \hbox{ and } D>B\,.\label{ManinR}\end{align}

The Manin relations define an ideal in the free algebra
$$\C_q\langle \hat g_{AB}\rangle,\qquad A,B=1, \dots, 4\,,$$ that we denote as $\cI_M$. The quotient algebra
$$M_q(4)=\C_q\langle \hat g_{AB}\rangle/\cI_M \qquad A,B=1, \dots, 4\,,$$ is the {\it quantum matrix algebra}. It is indeed a {\it bialgebra} with the coproduct defined on the generators as
\be \Delta \begin{pmatrix}
 \hat g_{11} &  \hat g_{12} &\hat g_{13} &  \hat g_{14}  \\
 \hat g_{21} &  \hat g_{22} & \hat g_{23} &  \hat g_{24} \\
 \hat g_{31} &  \hat g_{32} &  \hat g_{33} &  \hat g_{34}  \\
 \hat g_{41} &  \hat g_{42} &  \hat g_{43} &  \hat g_{44}  \\
\end{pmatrix}=\begin{pmatrix}
 \hat g_{11} &  \hat g_{12} &\hat g_{13} &  \hat g_{14}  \\
 \hat g_{21} &  \hat g_{22} & \hat g_{23} &  \hat g_{24} \\
 \hat g_{31} &  \hat g_{32} &  \hat g_{33} &  \hat g_{34}  \\
 \hat g_{41} &  \hat g_{42} &  \hat g_{43} &  \hat g_{44}  \\
\end{pmatrix}\otimes \begin{pmatrix}
 \hat g_{11} &  \hat g_{12} &\hat g_{13} &  \hat g_{14}  \\
 \hat g_{21} &  \hat g_{22} & \hat g_{23} &  \hat g_{24} \\
 \hat g_{31} &  \hat g_{32} &  \hat g_{33} &  \hat g_{34}  \\
 \hat g_{41} &  \hat g_{42} &  \hat g_{43} &  \hat g_{44}  \\
\end{pmatrix}\,,\label{qcoproduct}\ee
the matrix notation being evident.
\hfill$\blacksquare$

\end{definition}

The matrix bialgebra $M_q(n)$ can be defined for arbitrary $n\in \N$. We have  the following result \cite{ma}:

\begin{theorem}The map
\be\begin{CD} \cO_q(\cT)@>\tilde \Delta_q>>M_q(\C)\otimes\cO_q(\cT)\\
 \hat a^A@>>> \sum_B\hat g_{AB}\otimes \hat a^B\,\end{CD}\label{qCoaction}\ee
 is a {\it coaction} of $M_q(\C)$ on $\cO_q(\cT)=\C_q^4$. The Manin relations (\ref{ManinR}) are the necessary and sufficient condition.

\hfill$\blacksquare$
\end{theorem}

 Essentially, the Manin relations are  the commutation relations that the $\hat g_{AB}$'s have to satisfy in order to preserve the commutation relations (\ref{quantumtwistorrel}). For $q=1$ one recovers (\ref{coactionGL}).

\begin{definition} Given an $n\times n$ quantum matrix $M$, its {\it quantum determinant}  is defined as
$$ {\det}_q \,M=\sum_{\sigma\in S_n}(-q)^{-l(\sigma)}M_{n\sigma(n)}\cdots M_{1\sigma(1)}.$$
\hfill $\blacksquare$\end{definition}

We are ready now for the definition of the quantum group.

\begin{definition}

The  {\it quantum group } $\rSL_q(4,\C)$ is the free associative algebra over $\C_q$ with generators $\hat  g_{AB}$, $A,B=1,\dots, 4$ satisfying the Manin relations (\ref{ManinR})
 and the condition on the quantum determinant
\be {\det}_q \,\hat g=\sum_{\sigma\in S_4}(-q)^{-l(\sigma)}\hat g_{4\sigma(4)}\cdots \hat g_{1\sigma(1)}=1.\label{qdet}\ee
If we denote by $\cI_{\rSL_q(4,\C)}$ the ideal generated by (\ref{ManinR}) and (\ref{qdet}), then the algebra
$$\rSL_q(4,\C)=\C_q\langle  \hat  g_{AB} \rangle/\cI_{\rSL_q(4,\C)}$$
is a  quantum group \footnote{This is the standard notation instead of the more involved $\cO_q\big( \rSL(4,\C)\big)$.}.  This algebra is a deformation of $\cO\big( \rSL(4,\C)\big)$ as a Hopf algebra. The coproduct is given by (\ref{qcoproduct})and the antipode is a generalization of the formula (\ref{antipode})
$$S_q(\hat g_{AB})=(-q)^{B-A}M^q_{BA}\,,$$
where $M^q_{BA}$ is the corresponding quantum minor. One can see that   $S^2\neq \id$, contrary to what happens in the commutative case.  $\rSL_q(4,\C)$ is a non commutative, non cocommutative Hopf algebra.
\hfill$\blacksquare$
\end{definition}

\begin{definition}

The quantum Poincar\'{e} group times dilations, denoted as  $\cO_q(P)$,   is  the subalgebra of $\rSL_q(4,\C)$ generated by
\be \hat g=\begin{pmatrix}
 \hat g_{11} &  \hat g_{12} &\vline& 0 & 0  \\
 \hat g_{21} &  \hat g_{22} &\vline& 0 & 0  \\\hline
 \hat g_{31} &  \hat g_{32} &\vline&  \hat g_{33} &  \hat g_{34}  \\
 \hat g_{41} &  \hat g_{42} &\vline&  \hat g_{43} &  \hat g_{44}  \\
\end{pmatrix}.\label{qpoincare}
\ee
\hfill$\blacksquare$
\end{definition}

It will be convenient to use the alternative generators
$$ \hat g=\begin{pmatrix} \hat x&0\\ \hat T \hat x& \hat y\\\end{pmatrix}\,.$$ We can work out the form for $\hat T$ in terms of the old generators. In order to do so, we introduce the notation
$$\hat D^{KL}_{IJ} =\hat g_{IK}\hat g_{JL}-q^{-1}\hat g_{IL}\hat g_{JK},$$ (that is, they are $2\times 2$ quantum determinants). For simplicity, we will write
$$\hat D_{IJ}^{12}\equiv \hat D_{IJ}.$$
The condition  (\ref{qdet}) on the quantum determinant implies that $\det_q \hx=\hat D_{12}$ and $ \det_q \hy=D^{34}_{34}$ are invertible and
$${\det}_q \hx\cdot \,{\det}_q \hy=1.$$

The generators $\hT$ can now be computed explicitly:
$$ \hat T=\begin{pmatrix}-q^{-1}\hat D_{23}\det_q \hy& \hat D_{13}\det_q \hy\\-q^{-1}\hat D_{24}\det_q \hy&\hat D_{14}\det_q \hy\end{pmatrix}\,.$$

We give the commutation relations in the following proposition.

\begin{proposition}The commutation relations among the generators $\hat x_{ij},\, \hat y_{ab},\, \hat T_{ai}$ of  $\cO_q(P)$ are as follow:
\begin{align}
&\hat x_{11}\hat x_{12}=q^{-1}\hat x_{12}\hat x_{11}, &&
\hat x_{11}\hat x_{21}=q^{-1}\hat x_{21}\hat x_{11},\nonumber\\&
\hat x_{11}\hat x_{22}=\hat x_{22}\hat x_{11}+(q^{-1}-q)\hat x_{21}\hat x_{12}, &&
\hat x_{12}\hat x_{21}=\hat x_{21}\hat x_{12},\nonumber\\
&\hat x_{12}\hat x_{22}=q^{-1}\hat x_{22}\hat x_{12}, &&
\hat x_{21}\hat x_{22}=q^{-1}\hat x_{22}\hat x_{21}\,,\label{commx}
\end{align}
\begin{align}
&\hat y_{33}\hat y_{34}=q^{-1}\hat y_{34}\hat y_{33}, &&\hat y_{33}\hat y_{43}=q^{-1}\hat y_{43}\hat y_{33},\nonumber\\
&\hat y_{33}\hat y_{44}=\hat y_{44}\hat y_{33}+(q^{-1}-q)\hat y_{43}\hat y_{34}, &&
\hat y_{34}\hat y_{43}=\hat y_{43}\hat y_{34},\nonumber\\
&\hat y_{34}\hat y_{44}=q^{-1}\hat y_{44}\hat y_{34}, &&
\hat y_{43}\hat y_{44}=q^{-1}\hat y_{44}\hat y_{43}\,,\label{commy}
\end{align}
\begin{align}
&\hat T_{42} \hat T_{41}  =  q^{-1} \hat T_{41} \hat T_{42}, &&
\hat T_{31} \hat T_{41}  =  q^{-1} \hat T_{41} \hat T_{31}, \nonumber\\
& \hat T_{32} \hat T_{41}  =   \hat T_{41} \hat T_{32} + (q^{-1} -q ) \hat T_{42} \hat T_{31},\qquad
&& \hat T_{31} \hat T_{42}  =  \hat T_{42} \hat T_{31},\nonumber \\
& \hat T_{32} \hat T_{42}  =  q^{-1} \hat T_{42} \hat T_{32},  &&
 \hat T_{32} \hat T_{31}  =  q^{-1} \hat T_{31} \hat T_{32}\,.\label{commT} \end{align}

\noindent  and for $i=1,2$, $a=3,4$

 \begin{align}
 &\hx_{1i}\hT_{32}=\hT_{32}\hx_{1i}, &&\hx_{1i}\hT_{42}=\hT_{42}\hx_{1i}, &&
\hx_{1i}\hT_{31}=q^{-1}\hT_{31}\hx_{1i}, \nonumber\\&\hx_{1i}\hT_{41}=q^{-1}\hT_{41}\hx_{1i},  &&
\hx_{2i}\hT_{31}=\hT_{31}\hx_{2i},  &&\hx_{2i}\hT_{41}=\hT_{41}\hx_{2i},\nonumber\\
&\hx_{21}\hT_{a2}=q^{-1}\hT_{a2}\hx_{21}+q(q^{-1}-q)\hx_{11}\hT_{a1}&&\nonumber\\
&\hx_{22}\hT_{a2}=q^{-1}\hT_{a2}\hx_{22}+q(q^{-1}-q)\hx_{12}\hT_{a1} &&\label{commxt}
\end{align}

\begin{align}
&\hy_{33}\hT_{3a}=q\hT_{3a}\hy_{33}, && \hy_{34}\hT_{3a}=q\hT_{3a}\hy_{34},
&&\hy_{43}\hT_{4a}=q\hT_{4a}\hy_{43}, && \hy_{44}\hT_{4a}=q\hT_{4a}\hy_{44}, \nonumber\\
&\hy_{33}\hT_{4a}=\hT_{4a}\hy_{33}, && \hy_{34}\hT_{4a}=\hT_{4a}\hy_{34},
&&\hy_{43}\hT_{3a}=\hT_{3a}\hy_{43},  && \hy_{44}\hT_{3a}=\hT_{3a}\hy_{44}\,. \label{commyt}
\end{align}
\end{proposition}

{\sl Proof}.
This can be checked by direct computation \cite{cfl1,cfl2}.
\hfill$\blacksquare$

\medskip

 If we denote by $\cI_{P}$ the ideal generated by the relations  (\ref{commx}, \ref{commy}, \ref{commT}, \ref{commxt}, \ref{commyt}), then \be\cO_q(P)=\C_q\langle\hat x_{ij},\hat y_{ab}, \hat T_{ai}\rangle/(\cI_{P}, \; {\det}_q \hx \cdot {\det}_q \hy-1)\,.\label{presentation oqpl}\ee This is a Hopf subalgebra of $\rSL_q(4,\C)$.
The coproduct and the antipode are inherited form the ones in $\rSL_q(4,\C)$. It is instructive to compute the quantum antipode in terms of the variables $\hx, \hy, \hT$.
The coproduct is formally
 as in (\ref{coproductPl}), while for the antipode one has to replace the minors by quantum minors. Explicitly,
\begin{align*}
&S(\hat x)
={\det}_q \hy\begin{pmatrix} \hat x_{22} & -q\hat x_{12} \\ \\-q^{-1}\hat x_{21} & \hat x_{11} \end{pmatrix},\\\\
&S(\hat y)={\det}_q\hx\begin{pmatrix} y_{44} & -q\hat y_{34} \\\\ -q^{-1}\hat y_{43} & \hat y_{33} \end{pmatrix}, \\\\
&S(\hat T)=-S(\hat y)\hat T\hat x\,.\end{align*}

\bigskip

We are ready now to give a definition of the quantum Minkowski space mimicking (\ref{minkowskicoord}).

\begin{definition}\label{cqms}

The {\it complexified quantum Minkowski space} is the free algebra in four generators
$$
\hv_{41} , \hv_{42} , \hv_{31} \, \hbox{ and } \, \hv_{32}\,,
$$
satisfying the relations
\begin{align}
&\hv_{42} \hv_{41}  =  q^{-1} \hv_{41} \hv_{42}, \nonumber\\
&\hv_{31} \hv_{41}  =  q^{-1} \hv_{41} \hv_{31}, \nonumber\\
& \hv_{32} \hv_{41}  =   \hv_{41} \hv_{32} + (q^{-1} -q ) \hv_{42} \hv_{31},\nonumber\\
& \hv_{31} \hv_{42}  =  \hv_{42} \hv_{31}, \nonumber\\
& \hv_{32} \hv_{42}  =  q^{-1} \hv_{42} \hv_{32}, \nonumber\\
& \hv_{32} \hv_{31}  =  q^{-1} \hv_{31} \hv_{32}\,. \label{Relations QM}
\end{align}
Formally, these relations are the same as (\ref{commT}).

This algebra will be denoted as $\cO_q(\M)$. If we denote the ideal (\ref{Relations QM}) by $\cI_{\M_q}$, then we have that
$$\cO_q(\M)\equiv \C_q\langle \hv_{41}, \hv_{42}, \hv_{31}, \hv_{32}\rangle/\cI_{\M_q}\,.$$
\hfill$\blacksquare$
\end{definition}

It is not difficult to see that $\cO_q(\M)$ is isomorphic to the algebra of quantum matrices $M_q(2)$ (as defined for example in Ref. \cite{ka} or Ref. \cite{ma}). The correspondence $M_q(2)\rightarrow \cO_q(\M)$ is given in terms of the respective generators:
$$\begin{pmatrix}\hat{a}_{11}& \hat{a}_{12}\\ \hat{a}_{21}& \hat{a}_{22}\end{pmatrix}\rightleftarrows\begin{pmatrix}  \hv_{32}& \hv_{31}\\ \hv_{42}& \hv_{41}\end{pmatrix}.$$ Using this correspondence, one can check that the relations (\ref{Relations QM}) become the relations satisfied by the generators of the quantum matrices  $M_q(2)$.

One can check the following crucial fact:

\begin{proposition}There is a coaction of $\cO_q(P)$ on $\cO_q(\M)$, which on the generators has the same form as (\ref{coactionPl}).
\hfill$\blacksquare$
\end{proposition}
This justifies Definition \ref{cqms}.

At this stage, we have lost the interpretation in terms of functions over the Minkowski space. This will be recovered with the star product.

\section{Algebraic star product on  Minkowski space} \label{sec:Algebraic}

We consider now the algebra of the classical Minkowski space with the scalars extended to the ring $\C_q$
$$\cO(\rM)[q,q^{-1}]\equiv\C_q[t_{41}, t_{42},t_{31}, t_{32}].$$

\begin{proposition}\label{moduleiso} There is an isomorphism
 $\cO(\rM)[q,q^{-1}]\approx \cO_q(\M)$
 as modules over $\C_q$. In fact, the map
\be\begin{CD}\C_q[t_{41}, t_{42},t_{31}, t_{32}]@> Q_\M>>\cO_q(\M)\\t_{41}^a t_{42}^b t_{31}^c  t_{32}^d @>>> \hv_{41}^a \hv_{42}^b  \hv_{31}^c  \hv_{32}^d\end{CD}\label{QM}\ee is a module isomorphism (so it has an inverse).
\end{proposition}

{\sl Proof.}See Ref. \cite{ma}.\hfill$\blacksquare$

\medskip

  A map like (\ref{QM}) is called an {\it ordering rule} or {\it quantization map}. In particular, Proposition \ref{moduleiso} is telling us that   $\cO_q(\M)$ is a free module over $\C_q$, with basis the set of standard monomials.

We can  pull back the product on $\cO_q(\M)$  to  $\cO(\rM)[q,q^{-1}]$.

\begin{definition} The {\it star product} on $\cO(\rM)[q,q^{-1}]$ is defined  as
 \be f\star g=Q_\rM^{-1}\bigl(Q_{\rM}(f)Q_\rM(g)\bigr),\qquad f,g\in \cO(\rM)[q,q^{-1}].\label{starprodM}\ee\hfill$\blacksquare$
 \end{definition}

 By construction, the star product satisfies associativity. The algebra $(\cO(\rM)[q,q^{-1}], \,\star\,)$ is then isomorphic to $\cO_q(\M)$. Working on $\cO(\rM)[q,q^{-1}]$ has the advantage of working with classical objects (the polynomials), were one has substituted the standard pointwise product by the noncommutative star product. This is important for the physical applications. Moreover, we can study if this star product has an extension  to all the $C^\infty$ functions, and if the extension is  differential. If so,  Kontsevich's theory \cite{ko} would then be relevant.

 We want to obtain a formula for the star product.
We begin by computing some  auxiliary relations

\begin{lemma} The following commutation rules are satisfied in $\cO_q(\M)$:
\begin{align*}
&\hv_{42}^m \hv_{41}^n = q^{-mn} \hv_{41}^n \hv_{42}^m, \\
& \hv_{31}^m \hv_{41}^n = q^{-mn} \hv_{41}^n \hv_{31}^m, \\
& \hv_{31}^m \hv_{42}^n = \hv_{42}^n \hv_{31}^m, \\
& \hv_{32}^m \hv_{42}^n = q^{-mn} \hv_{42}^n  \hv_{32}^m, \\
& \hv_{32}^m \hv_{31}^n = q^{-mn} \hv_{31}^n \hv_{32}^m,
\end{align*}
and
$$
\hv_{32}^m \hv_{41}^n = \hv_{41}^n \hv_{32}^m + \sum_{ k = 1}^{\mu} F_k(q, m,n) \hv_{41}^{n-k} \hv_{42}^k \hv_{31}^k \hv_{32}^{m-k},
$$
where
$
\mu = \mathrm{min}(m,n)
$
\be
F_k(q,m,n) = \beta_k(q,m) \prod_{l = 0 }^{k-1} F(q , n - l )\qquad \hbox{with}\qquad
F(q,n) = \left(\frac{1}{q^{2 n - 1}} -q\right)\label{efe}\ee
and $ \beta_k(q,m) $ defined by the recursive relation
$$
\beta_0(q,m)= \beta_m(q,m) = 1, \quad  \hbox{and}\quad \beta_k(q,m+1) = \beta_{k-1}(q,m) + \beta_k(q,m) q^{-2 k}.
$$
Moreover, $\beta_k(q,m)=0$ if $k<0$ or if $k>m$.

\end{lemma}

{\sl Proof}. The proof is just a (lengthy) computation. \hfill$\blacksquare$

\medskip
Using the above  relations, we obtain the final result:

\begin{theorem}\label{explicitstar} The star product defined in Definition \ref{starprodM} is given on two arbitrary monomials as
\begin{align}
&(t_{41}^a t_{42}^b t_{31}^c t_{32}^d )\star (t_{41}^m t_{42}^n t_{31}^p t_{32}^r)  =  q^{-mc-mb-nd-dp} t_{41}^{a+m} t_{42}^{b+n} t_{31}^{c+p} t_{32}^{d+r} \; + \nonumber\\ & \sum_{k = 1}^{\mu=min(d,m)} q^{-(m-k)c -(m-k)b - n(d-k)-p(d-k)}  F_k(q,d,m) \; t_{41}^{a+m-k} t_{42}^{b+k+n} t_{31}^{c+k+p} t_{32}^{d-k+r}\label{starpoly}
\end{align}

\hfill$\blacksquare$

\end{theorem}

\section{Differential star product on the big cell}	\label{sec:Differential}
In order to compare the algebraic star product obtained above with the differential star product approach we consider a change in the parameter, $q=\exp h$. The classic limit is obtained as $h\rightarrow 0$. We will expand (\ref{starpoly}) in powers of $h$ and we will show that each term can be written as a bidifferential operator. Then the extension of the star product to $C^\infty$ functions is unique.

\begin{theorem} We consider $$q=\re^h=\sum \frac {h^n}{n!}\,,$$ and we expand the star product of Theorem \ref{explicitstar} in powers of $h$. Then, at each order in $h$, one can find a bidifferential operator that reproduces the result of the formula (\ref{starpoly}). \hfill$\blacksquare$

\end{theorem}

We devote the rest of the section to the proof of this theorem.

\subsection{Explicit computation up to order 2}

We first take up the explicit computation of the bidifferential operators  up to order 2. Then we will argue that a differential operator can be found at each order.

\bigskip

We rewrite (\ref{starpoly}) as
$$f\star g= fg+ \sum_{j=1}^\infty h^j C_j(f, g),$$
with $$
f = t_{41}^a t_{42}^b t_{31}^c t_{32}^d, \qquad  g = t_{41}^m t_{42}^n t_{31}^p t_{32}^r.$$
At order 0 in $h$ we recover the commutative product. At order $n$ in $h$ we have contributions from each of the terms with different $k$ in (\ref{starpoly}).
$$C_n(f,g)= \sum_{k=0}^{\mu=min(d,m)} C_n^{(k)}(f,g),$$ (the terms with $k=0$ come from the first term in (\ref{starpoly})).

Let us compute each of the contributions $C_1^{(k)}$:
\begin{itemize}
\item $k = 0$. We have

$$C_1^{(0)}=(-mc-mb-nd-dp)\; t_{41}^{a+m}t_{42}^{b+n}t_{31}^{c+p}t_{32}^{d+r }.$$
It is easy to see that this is reproduced by the  bidifferential operator
\begin{displaymath}
C_1^{(0)}(f,g)=-( t_{41} t_{31} \partial_{31} f \partial_{41} g + t_{42} t_{41} \partial_{42} f \partial_{41} g + t_{32} t_{42} \partial_{32} f \partial_{42} g  +   t_{32} t_{31} \partial_{32} f \partial_{31} g).
\end{displaymath}
We will denote the bidifferential operators by means of the tensor product (as it is customary). For example
\begin{displaymath}
C_1^{(0)}= -(t_{41} t_{31} \partial_{31} \otimes \partial_{41}  + t_{42} t_{41} \partial_{42} \otimes \partial_{41}  + t_{32} t_{42} \partial_{32} \otimes \partial_{42} +   t_{32} t_{31} \partial_{32} \otimes \partial_{31}),
\end{displaymath}
so
$$C_1^{(0)}(f,g)=C_1^{(0)}(f\otimes g).$$

\item $ k = 1 $. Let us first compute the factor $ F_1(q,d,m)=\beta_1(q,d) F(q,m) $. First, notice that
    \begin{align*} \beta_1(q,d) =& 1 + q^{-2} + q^{-4} + \dots + q^{-2(d-1)} = \frac{\re^{-2dh}-1}{\re^{-2h}-1}=\\&d -d(d-1)h+\frac 13d((1-3d+2d^2)h^2+\cO(h^3),
\end{align*} and that
    $$F(q,n) = -2 n h+2n(n-1) h^2 +\cO(h^3),$$
so up to order $h^2$ we have  $$ \beta_1(q,d) F(q,m) = -2 m  d h+2md(d+m-2) h^2 +\cO(h^3).$$ Finally,
 the contribution  of the $k=1$ term  to  $C_1$ is
 $$C_1^{(1)}(f,g)=-2 m  d t_{41}^{a+m-1} \;t_{42}^{b+n+1} t_{31}^{c+p+1} t_{32}^{d+r-1}.$$

This is reproduced by the bidifferential operator
$$C_1^{(1)}=
-2t_{42} t_{31} \partial_{32} \otimes \partial_{41} .
$$

\item $ k \geq 2 $ We have the factor
$$
 \beta_k(q,d) F(q , m) F(q,m-1) \cdots F(q,m-k)= \cO(h^k),
 $$ so the terms with $ k \geq 2 $ do not contribute $C_1$.
\end{itemize}

Summarizing,
\begin{align}
C_1  =& C_1^{(0)}+C_1^{(1)}=  - ( t_{41} t_{31} \partial_{31} \otimes \partial_{41}  + t_{42} t_{41} \partial_{42} \otimes \partial_{41}  +\nonumber\\ & t_{32} t_{42} \partial_{32} \otimes \partial_{42} +     t_{32} t_{31} \partial_{32} \otimes \partial_{31}  + 2 t_{42} t_{31} \partial_{32} \otimes \partial_{41}  ),\label{C1}
\end{align}
so $C_1$ is extended to the  $C^\infty$ functions. If we antisymmetrize $C_1$ we obtain a Poisson bracket
\begin{align}
\{f,g \}=  &t_{41} t_{31} (\partial_{41}f \partial_{31}g-\partial_{41}g \partial_{31}f)  + t_{42} t_{41} (\partial_{41} f \partial_{42} g-
\partial_{41} g \partial_{42} f)+\nonumber\\& t_{32} t_{42} (\partial_{42}f  \partial_{32} g-\partial_{42}g  \partial_{32} f)+
    t_{32} t_{31} (\partial_{31} f\partial_{32}g-\partial_{31} g\partial_{32}f)  + \nonumber\\&2 t_{42} t_{31}( \partial_{41}f \partial_{32}g-\partial_{41}g  \partial_{32}f )
\label{pbt}
\end{align}
We can express the Poisson bracket in terms of the usual variables in Minkowski space. Using (\ref{Pauli}), the change of coordinates is
$$\begin{pmatrix}t_{31}&t_{32}\\t_{41}&t_{42}\end{pmatrix}= x^\mu \sigma_\mu=\begin{pmatrix} x^0 + x^3 & x^1 - \ri x^2 \\ x^1 + \ri x^2 & x^0 - x^3 \end{pmatrix},$$ and the inverse change is
$$x^{0}=\frac{1}{2}(t_{31}+t_{42}),\quad
x^{1}=\frac{1}{2}(t_{32}+t_{41}),\quad
x^{2}=\frac{i}{2}(t_{32}-t_{41}),\quad
x^{3}=\frac{1}{2}(t_{31}-t_{42}).$$
In these variables the Poisson bracket is
\begin{align}
\{f,g \}= &\ri\Big((x^{0})^{2}-(x^{3})^{2})(\partial_{1}f\partial_{2}g -\partial_{1}g\partial_{2}f)  + x^{0}x^{1}(\partial_{0}f\partial_{2}g -\partial_{0}g\partial_{2}f)-\nonumber \\&
x^{0}x^{2}(\partial_{0}f\partial_{1}g
-\partial_{0}g\partial_{1}f)-x^{1}x^{3}(\partial_{2}f\partial_{3}g -\partial_{2}g\partial_{3}f)
+\nonumber\\&x^{2}x^{3}(\partial_{1}f\partial_{3}g -\partial_{1}g\partial_{3}f)\Big)
\label{pb2}
\end{align}

\bigskip

We now compute  the term $C_2$. We sum the contributions to the order $h^2$ of each term in (\ref{starpoly})

\begin{itemize}
\item $ k = 0 $. The contribution to the order $h^2$ is
$$ C_2^{(0)}=\frac{1}{2}(mc+mb+nd+dp)^2 \;t_{41}^{a+m} t_{42}^{b+n} t_{31}^{c+p} t_{32}^{d+r}. $$
This is reproduced by
\begin{align*}
C_2^{(0)}=&\frac{1}{2} t_{31} t_{41}\;\partial_{31} (  t_{31} \partial_{31}  )\otimes\partial_{41} (  t_{41} \partial_{41} ) + t_{42} t_{31} t_{41} \; \partial_{42} \partial_{31} \otimes\partial_{41}(  t_{41} \partial_{41} ) +\\&
 t_{31} t_{32} t_{41} t_{42}\; \partial_{31} \partial_{32} \otimes\partial_{41} \partial_{42}  + t_{31}^2 t_{32} t_{41} \;\partial_{31 } \partial_{32} \otimes \partial_{41} \partial_{31} +
\\ &  \frac{1}{2}t_{42} t_{41} \;\partial_{42} ( t_{42} \partial_{42} ) \otimes\partial_{41}( t_{41} \partial_{41} )  +   t_{41} t_{42}^2 t_{32} \partial_{42 } \partial_{32} \otimes \partial_{41} \partial_{42} +
\\  &  t_{41} t_{42} t_{31} t_{32} \partial_{42 } \partial_{32} \otimes \partial_{41} \partial_{31}  + \frac 1 2t_{32} t_{42} t_{31}  \partial_{32}(  t_{32} \partial_{32}  ) \otimes \partial_{42} \partial_{31} +
\\ &  \frac 1 2t_{32} t_{31}\;\partial_{32}(  t_{32} \partial_{32}  ) \otimes\partial_{31}  ( t_{31} \partial_{31} )   +  t_{32} t_{42} t_{31}\;\partial_{32}  ( t_{32} \partial_{32}) \otimes\partial_{42} (t_{42} \partial_{42} ).
\end{align*}

\item $ k = 1 $. We have that
 $$F_1(q,d,m)=\beta_1(q,d)F(q, m).$$ Expanding both factors  we have
 \begin{align*}\beta_1(q,d) & = d - d(d-1) h + O(h^2),\\
 F(q,m) & = -2mh-2m(1-m) h^2 + O(h^3),\end{align*} so we get
$$
\beta_1(q,d) F(q,m) \approx  -2mdh +2md(( m-1 ) + (d -1 ))h^2,
$$
and the contribution to order $h^2$ is
\begin{align*}&h^2\,\bigg(2md\Big((m-1) + (d-1) + (m-1)c+(m-1)b+n(d-1)+p(d-1)\Big)\bigg)\,\cdot\\&\cdot\,t_{41}^{a+m-1} t_{42}^{b+n+1} t_{31}^{c+p+1} t_{32}^{d+r-1}.\end{align*}

 We reproduce that result with
\begin{align*}
C_2^{(1)}=&2 t_{32} t_{42} t_{31} \partial_{32}^2 \otimes \partial_{41}  + 2 t_{31} t_{42} t_{41} \partial_{32} \otimes \partial_{41}^2  - 2 t_{31} t_{42}^2 t_{41} \partial_{42} \partial_{32} \otimes \partial_{41}^2 +
\\ &  2 t_{42} t_{31}^2 t_{41} \partial_{31} \partial_{32} \otimes \partial_{41}^2  +2 t_{31} t_{42}^2 t_{32} \partial_{32}^2 \otimes \partial_{41} \partial_{42}  +2 t_{42} t_{31}^2 t_{32} \partial_{32}^2 \otimes \partial_{41} \partial_{31}
\end{align*}

\item $ k = 2$. One can show  that
$$ \beta_2(q,d) = \frac{d(d-1)}{2} + O(h),$$ so $$ \beta_2(q,d) F(q,m) F(q,m-1) \approx 2 d (d-1) m (m-1) h^2,$$ and the contribution of this term to the order $h^2$ is
$$
h^2\, 2 d (d-1) m (m-1)  \; t_{41}^{a+m-2} t_{42}^{b+n+2} t_{31}^{c+p+2} t_{32}^{d+r-2}.
$$
This is given by
$$
 C_2^{(2)}=2 t_{42}^2 t_{31}^2 \partial_{32}^2 \otimes \partial_{41}^2.
$$

\end{itemize}
Summarizing we get
\begin{align*}
C_2 =&\frac{1}{2} t_{31} t_{41} \;\partial_{31}(   t_{31} \partial_{31}  )\otimes  \partial_{41}( t_{41} \partial_{41} ) + t_{42} t_{31} t_{41} \partial_{42} \partial_{31}\otimes  \partial_{41}( t_{41} \partial_{41} )+
\\ &
t_{31} t_{32} t_{41} t_{42} \partial_{31} \partial_{32} \otimes \partial_{41} \partial_{42} + t_{31}^2 t_{32} t_{41} \partial_{31 } \partial_{32} \otimes \partial_{41} \partial_{31}+
\\&
\frac 1 2{t_{42} t_{41}}\;\partial_{42} (t_{42} \partial_{42} )\otimes \partial_{41}( t_{41} \partial_{41})  +
\\&
 t_{41} t_{42}^2 t_{32} \partial_{42 } \partial_{32} \otimes\partial_{41} \partial_{42} t_{41} t_{42} t_{31} t_{32} \partial_{42 } \partial_{32} \otimes \partial_{41} \partial_{31}  +
 \\ &
 \frac 1 2 t_{32} t_{42}  \partial_{32}( t_{32} \partial_{32}  )\otimes \partial_{42} \partial_{31}+
  \frac 1 2{t_{32} t_{31}} \partial_{32} (t_{32} \partial_{32})  \otimes \partial_{31}( t_{31} \partial_{31} )   +
 \\\\ &
   t_{32} t_{42} t_{31} \partial_{32}(  t_{32} \partial_{32} )\otimes \partial_{42} \partial_{31}+ 2 t_{42}^2 t_{31}^2 \partial_{32}^2 \otimes \partial_{41}^2  +
  \\\\&
  2 t_{32} t_{42} t_{31} \partial_{32}^2 \otimes \partial_{41}  + 2 t_{31} t_{42} t_{41} \partial_{32} \otimes \partial_{41}^2- 2 t_{42} t_{31}^2 t_{41} \partial_{31} \partial_{32} \otimes \partial_{41}^2  -
  \\\\ &
  2 t_{31} t_{42}^2 t_{32} \partial_{32}^2 \otimes \partial_{41} \partial_{42}  -2 t_{42} t_{31}^2 t_{32} \partial_{32}^2 \otimes \partial_{41} \partial_{31} -  2 t_{31} t_{42}^2 t_{41} \partial_{42} \partial_{32} \otimes \partial_{41}^2.
\end{align*}

\subsection{Differentiability at arbitrary order}

We are going to prove now the differentiability of the star product. We keep in mind the expression (\ref{starpoly}), which has to be expanded in $h$. Our goal will be to show that,
 at each order, it can be reproduced by a bidifferential operator with no dependence on the exponents $a, b,c, d, m, n, p, r$.

Let us first argue on a polynomial function of one variable, say $x$. For example, we have
$$
m\;x^{m-1}=\partial_x \big(x^{m}\big).
$$
More generally, we have
\begin{align}
&m^b\; x^{m}=  (x\partial_x)^{b}\big(x^{m}\big) \qquad \hbox{ and }\nonumber\\&m^b (m-1)^c \cdots (m -k +1)^d \; x^{m-k}  =\partial_x (x \partial_x)^{d-1} \hdots  \partial_x (x \partial_x)^{c-1}   \partial_x(x \partial_x)^{b-1}\big(x^{m}\big).\\&\label{difop}
\end{align}
Notice that in the last formula, we have $ b,c,\dots ,d\geq 1$, otherwise the formula makes no sense. In fact, an arbitrary polynomial
$$p(x)=
\sum_{k\in \Z}f_k(m, x) x^{m-k},$$ is not generically obtainable from  $x^m$ by the application of a differential operator with coefficients that are independent of the exponents and polynomial in the variable $x$. One can try for example with $p(x)=x^{m-1}$. We then have that
$$x^{m-1}=\frac1m \partial_x (x^m),\qquad \hbox{or}\qquad x^{m-1}=\frac 1x x^m.$$
So the right combinations should appear in the coefficients in order to be reproduced by a differential operator with polynomial coefficients.

Let us see the  contribution  of  the terms with different $k$ in (\ref{starpoly}).
We start with the term $ k = 0 $. From
$$ q^{-mc-mb-nd-dp}\; t_{41}^{a+m} t_{42}^{b+n}t_{31}^{c+p} t_{32}^{d+r}$$ we  only get terms of the form
$$  b^{i_b} c^{i_c} d^{i_d}m^{i_m}n^{i_n}p^{i_p} \;t_{41}^{a+m} t_{42}^{b+n}t_{31}^{c+p} t_{32}^{d+r}.$$ Applying the rules (\ref{difop}), these terms can be easily reproduced by the bidifferential operators of the form
$$(t_{42}\partial_{42})^{i_b}(t_{31}\partial_{31})^{i_c}(t_{32}\partial_{32})^{i_d}\otimes
(t_{41}\partial_{41})^{i_m}(t_{42}\partial_{42})^{i_n}(t_{31}\partial_{31})^{i_p},
$$
applied to
$$t_{41}^{a} t_{42}^{b}t_{31}^{c} t_{32}^{d}\otimes t_{41}^{m} t_{42}^{n}t_{31}^{p} t_{32}^{r}.$$

We turn now to the more complicated case of   $ k \neq 0 $. We have to consider the two factors in (\ref{starpoly})
$$
q^{-(m-k)c -(m-k)b - n(d-k)-p(d-k)},\qquad \hbox{and }\quad    F_k(q,d, m).
$$
 Expanding both factors in powers of $h$ it is easy to see that the coefficients at each order are polynomials in $m, n, p, b, c, d, k$. What we have to check is that this polynomials have a form that can be reproduced with a bidifferential operator using (\ref{difop}).
 Let us start with
 $$F_k(q,d, m)=\beta_k(q,m)\prod_{l=0}^{k-1}F(q, m-l).$$
From the definition (\ref{efe}), we have that
 $
 F(q, j )|_{j = 0} = 0$, so
 $$ F(q,j ) = j G(q,j),  $$ with  $G(q,j)$  a series in $h$ with coefficients that are  polynomial in $j$. More generally, the product
   $$  L_k(q,m) = \prod_{l = 0 }^{k-1} F(q , m - l ) = m (m-1) (m-2) \cdots (m-k+1) L'(q,m).$$ The polynomials in $L'(q,m)$ are easily obtained with combinations of differential operators of the form
   $$\big(t_{41}\partial_{41}\big)^i(t_{41}^m).$$ The remaining factor $m (m-1) (m-2) \cdots (m-k+1) t_{41}^{m-k}$  is adjusted with the differential operator
   $$\partial_{41}^k (t_{41}^m)=m (m-1) (m-2) \cdots (m-k+1)\; t_{41}^{m-k}.$$

Let us work now with $ \beta_k(q,m)) $. We have that
 $$ \beta_k(q, m) = 0 \qquad  \hbox{ for } m< k, $$
 so
 $$
\beta_k(q, d ) = d ( d- 1) ( d - 2 ) \cdots (d -k +1 ) \beta'_k(q,d),
$$
with $ \beta_k'(q,d) $ a series in $h$ with coefficients that are polynomial in $d$. The differential operator that we need is of the form
$$\partial_{32}^k(\big(t_{32}\partial_{32})\big)^j \, (t_{32}^d)= d ( d- 1) ( d - 2 ) \cdots (d -k +1 )\, d^j\, t_{32}^{d-k}.$$

Finally, the factor $ q^{-(m-k)c -(m-k)b - n(d-k)-p(d-k)} $ introduces factors of the form
$$b^{i_b}c^{i_c}(d-k)^{i_d}(m-k)^{i_m}n^{i_n}p^{i_p}\; t_{41}^{a+m-k} t_{42}^{b+k+n} t_{31}^{c+k+p} t_{32}^{d-k+r},$$
which are  reproduced by
$$t_{42}^{k} t_{31}^{k} \big(t_{42}\partial_{42} \big)^{i_b}
\big(t_{31}\partial_{31} \big)^{i_c}\big(t_{32}\partial_{32} \big)^{i_d}\otimes
\big(t_{41}\partial_{41} \big)^{i_m}\big(t_{42}\partial_{42} \big)^{i_n}
\big(t_{31}\partial_{31} \big)^{i_p},$$
acting on
$$t_{41}^{a}t_{42}^{b}t_{31}^{c}t_{32}^{d-k}\otimes
t_{41}^{m-k}t_{42}^{n}t_{31}^{p}t_{32}^{r}.$$

This completes the proof of differentiability of the star product at arbitrary order.

\section{Poincar\'{e} coaction} \label{sec:Poincare}

 We would like to see how the coaction over the Minkowski space looks in terms of the star product, and if it is also differential. But in order to do so, we need first to have a star product on the group $\cO_q(P)$.

 \begin{theorem}
 The map
 \begin{align}\begin{CD}\cO(P)[q, q^{-1}]@>Q_G>>\cO_q(P)\\
\hy_{44}^a\hy_{43}^b\hy_{34}^c\hy_{33}^d \hx_{22}^e\hx_{21}^f\hx_{12}^g\hx_{11}^l \hT_{41}^m\hT_{42}^n\hT_{31}^p\hT_{32}^r@>>>
y_{44}^ay_{43}^by_{34}^cy_{33}^d x_{22}^ex_{21}^fx_{12}^gx_{11}^l T_{41}^mT_{42}^nT_{31}^pT_{32}^r\end{CD}\nonumber \\\label{QG}\end{align}
is a $\C_q$-module isomorphism. In particular,  $\cO_q(P)$ is a free module.

 \end{theorem}
First of all, we notice that the subalgebra generated by $\{\hat x_{ij}\}$  and $\{\hat y_{ab}\}$ are two copies of the algebra of $2\times 2$ quantum matrices, which commute among them. The maps to the standard quantum matrices  \cite{ka,ma} are this time

$$\begin{pmatrix}\hat{a}_{11}& \hat{a}_{12}\\ \hat{a}_{21}& \hat{a}_{22}\end{pmatrix}\rightleftarrows\begin{pmatrix}  \hx_{11}& \hx_{12}\\ \hx_{21}& \hx_{22}\end{pmatrix};\qquad\begin{pmatrix}\hat{a}_{11}& \hat{a}_{12}\\ \hat{a}_{21}& \hat{a}_{22}\end{pmatrix}\rightleftarrows\begin{pmatrix}  \hy_{33}& \hy_{34}\\ \hy_{43}& \hy_{44}\end{pmatrix},$$
as can be deduced from (\ref{commx}) and (\ref{commy}). One can chose the Manin order in each subset of variables,
$$
\hy_{44}<\hy_{43}<\hy_{34}<\hy_{33},\qquad \hx_{22}<\hx_{21}<\hx_{12}<\hx_{11}.
$$
With this one can construct a quantization map (given by the standard monomials basis) for the quantum Lorentz plus dilations group . We have now to include the translations to have the complete quantization map for the Poincar\'{e} group. It is clear that one can choose the Manin order also for the variables $\hT$, but, since these variables do not commute with the $\hat x$ and $\hat y$ we have to be careful in choosing a full ordering rule. This is a non trivial problem, but it can be solved. In Appendix \ref{diamond} we show that the ordering

$$\hy_{44}<\hy_{43}<\hy_{34}<\hy_{33}< \hx_{22}<\hx_{21}<\hx_{12}<\hx_{11}< \hT_{41}<\hT_{42}<\hT_{31}<\hT_{32}
$$
gives standard monomials that form a basis for the quantum Poincar\'{e} group $\cO_q(P)$.
As we did for the Minkowski space star product (\ref{starprodM}), we extend the scalars of the commutative algebra to $\C_q$ and   define a quantization map $Q_G$

\hfill$\blacksquare$

\medskip

\begin{definition} For  $f, g\in \cO(P)[q, q^{-1}]$,  the star product is defined as
$$f\star_G g= Q_G^{-1}(Q_G(f) \cdot Q_G(g ) ).$$
\hfill$\blacksquare$
\end{definition}

Let us now consider the coaction, formally as in (\ref{coactionPl}). Using both quantization maps ($Q_\M$ and $Q_G$)  we can define a star coaction:

\begin{proposition} The map
$$
\begin{CD} \cO(\M)[q,q^{-1}]@>\tilde\Delta_\star>> \rightarrow \cO(G)[q,q^{-1}] \otimes \cO(\rM)[q,q^{-1}]\\f@>>> Q_{G}^{-1} \otimes Q_{M}^{-1} ( \Delta ( Q_M ( f ) )\end{CD}
$$
has the compatibility property (see (\ref{compatibility}))
\be
\tilde\Delta_\star( f \star_M  g)  =  \tilde\Delta_\star (f )(\star_G\otimes \star_M) \tilde\Delta_\star ( g ),\qquad f,g\in \cO(\M)[q,q^{-1}].\label{deltastar}
\ee
so it defines a coaction of $(\cO(\M)[q,q^{-1}],\star_G)$ on $\cO(\M)[q,q^{-1}],\star_M)$.
\end{proposition}

{\sl Proof.} It follows from the definitions.

\hfill$\blacksquare$

\medskip

\subsection{The coaction as a differential operator}

We will restrict to the Lorentz group times dilations, that is, we will consider only the generators $x$ and $y$.

On the generators of Minkowski space the star coaction is simply
$$
\Delta_\star (  t_{ai}  ) = y_{ab}S(x_{ji})\otimes t_{bj},
$$
and, using the notation
$$t_{mi}^{\star a}=\underbrace{ t_{mi} \star_\M t_{mi} \star_\M \cdots\star_\M t_{mi} }_{a\, \mathrm{times}},$$ for an arbitrary standard monomial the coaction is expressed as
\begin{align*}
&\tilde\Delta_\star \left( t_{41}^a t_{42}^b t_{31}^c t_{32}^d \right)    = \tilde \Delta_{\star} (t_{41}^{\star a}\star_\M t_{42}^{\star b}\star_\M  t_{31}^{\star c}\star_\M  t_{32}^{\star d})=\\
&({\tilde\Delta_{\star} t_{41})}^{\star a}(\star_G\otimes\star_\M)({\tilde\Delta_{\star} t_{42})}^{\star b}
(\star_G\otimes\star_\M)({\tilde\Delta_{\star} t_{31})}^{\star c}
(\star_G\otimes\star_\M)({\tilde\Delta_{\star} t_{32})}^{\star d}.\end{align*}
We have used the exponent `$\star$' to indicate`$\star_\M$', `$\star_G$' or `$\star_{G\times \M}$' to simplify the notation. The meaning should be clear from the context.
Contracting with $ \mu_{G\times \M}$ (see (\ref{mu}) for the notation) we define
$$
\tau_{ij} \equiv \mu_{G\times \M}  \circ \tilde\Delta_{\star}( t_{ij} ) = y_{ab} t_{bj}S(x_{ji});
$$

Applying $ \mu_{G\times \M} $ to the coaction, we get

\be
\mu_{G\times \M} \circ \Delta_{\star}( t_{41}^{ a} t_{42}^{b} t_{31}^{c} t_{32}^{d} )    = \tau_{41}^{\star a}  \star_{G\times \M} \tau_{42}^{\star b}  \star_{G\times \M}  \tau_{31}^{\star c} \star_{G\times \M} \tau_{32}^{\star d}.\label{starcoaction}
\ee
Notice that in each $\tau$ there is a sum of terms with factors $ytS(x)$ that generically do not commute. So we need to work out the star products in the right hand side of (\ref{starcoaction}).

As we are going to see, the calculation is involved. We are going to make a change in the parameter $q=\exp h$ and expand the star product in  power series of $h$. At the end, we will compute only the first order term in $h$ of the star coaction.

The star product $\star_{G\times \M}$ is written, as usual,
$$
f_1 \star_{G\times \M}  f_2 = \sum_{m=0}^\infty h^m D_m ( f_1 , f_2 ), \qquad f_1, f_2 \in \cO(g\times\M)[[h]].
$$
For our purposes it will be enough to consider    $f_1$ and $f_2$ to be polynomials in $\tau$. The generators $x$, $y$ and $t$ commute among themselves, so the star product in $G\times \M$ can be computed reordering the generators in each group $x$, $y$, and $t$ in the Manin ordering. The result will be terms similar to the star product (\ref{starpoly}), and in particular, $D_1$ will contain 3 terms of the type $C_1$ (\ref{C1}), on for the variables $x$, another for the variables $y$ and another for the variables $t$. But $C_1$ is a bidifferential operator of order 1 in each of the arguments, so it satisfies the Leibnitz rule
$$
D_1( f_1 , f_2 u ) = D_1 (f_1 , f_2 ) u +D_1(f_1,u) f_2,\qquad u\in \cO(P)[q,q^{-1}]\,,
$$
then we have, for example,
 \be \label{leibniz}D_1( \tau_{ij} , \tau_{kl}^a) = a D_1 ( t_{ij} , t_{kl} )  t_{kl}^{a-1},\ee  which will be used in the following.

In general, we have
\begin{align*}
&\tau_{41}^{\star a} \star \tau_{42}^{\star b } \star \tau_{31}^{ \star c} \star \tau_{32}^{ \star d} = \sum_{I\in\cI} h^M D_{i_1} ( \tau_{41} , D_{i_2} ( \tau_{41}, \dots D_{i_{a-1}} ( \tau_{41} , D_{j_1} ( \tau_{42} ,
\\&
 D_{j_2} ( \tau_{42} , \dots D_{j_{b-1}} ( \tau_{42} , D_{l_1} ( \tau_{31} , D_{l_2} ( \tau_{31} , \dots D_{l_{D-1}} ( \tau_{31} , D_{m_1 } ( \tau_{32} ,
  \\&
  D_{m_2} ( \tau_{32} , \dots D_{m_{d-1}}( \tau_{32} , \tau_{32} ) \dots )
\end{align*}

Here $ M = i_{1} + \hdots + i_{a} + j_1 + \hdots + j_b + l_1 + \hdots + l_c + m_1 + \hdots m_d $ and we sum over all the multiindices
 $$ I = \left( i_1 , \hdots , i_{a-1} , j_1 , \hdots , j_{b-1} , l_1 , \hdots , l_{c-1} , m_1 , \hdots , m_{d-1} \right).$$ We are interested in the first order in $h$, so $ M = 1 $. This means that for any term in the sum we have only one $ D_1 $ operator (the others are $ D_0$, which is just the standard product of both arguments). So we have the sum

\begin{align*}
 \sum_{k}  &\Big(\tau_{41}^{k}D_{1} ( \tau_{41}, \tau_{41}^{  a-k-1} \tau_{42}^{   b} \tau_{31}^{  c} \tau_{32}^{    d } ) +  \tau_{41}^{   a} \tau_{42}^{    k} D_1 ( \tau_{42}^ , \tau_{42}^{b-k-1} \tau_{31}^{  c} \tau_{32}^{  d} ) +\\  & \tau_{41}^{  a} \tau_{42}^{  b} \tau_{31}^{  k} D_1 ( \tau_{31}^ , \tau_{31}^{  c-k-1} \tau_{32}^{  d} )+   \tau_{41}^{  a} \tau_{42}^{  b} \tau_{31}^{  c} \tau_{32}^{  k} D_1 ( \tau_{32}, \tau_{32}^{  d-k-1} )\Big).
\end{align*}
using (\ref{leibniz}) we get
\begin{align*}
 \sum_{k = 1}^{a-1}  k  \tau_{41}^{  a-2} \tau_{42}^{  b} \tau_{31}^{  c} \tau_{32}^{   d} D_1 ( \tau_{41}  , \tau_{41}  )  + \sum_{k = 1 }^a b \tau_{41}^{  a-1} \tau_{42}^{  b-1} \tau_{31}^{  c} \tau_{32}^{  d} D_1 ( \tau_{41} , \tau_{42})   +
\\  \sum_{k= 1 }^a c \tau_{41}^{  a-1} \tau_{42}^{ b} \tau_{31}^{ c-1} \tau_{32}^{ d} D_1 ( \tau_{41} , \tau_{31} )  +   \sum_{k = 1 }^a d \tau_{41}^{a-1} \tau_{42}^{ b} \tau_{31}^{ c} \tau_{32}^{ d-1} D_1 ( \tau_{41} , \tau_{32} )   +
\\  \sum_{k = 1 }^{b-1} k \tau_{41}^{ a} \tau_{42}^{ b-2} \tau_{31}^{ c} \tau_{32}^{ d} D_1 ( \tau_{42} , \tau_{42}) + \sum_{k = 1 }^b c \tau_{41}^{ a} \tau_{42}^{ b-1} \tau_{31}^{ c-1} \tau_{32}^{ d} D_1 ( \tau_{42} \tau_{31} )  +
\\ \sum_{k = 1 }^b d \tau_{41}^{ a} \tau_{42}^{ b-1} \tau_{31}^{ c} \tau_{32}^{ d-1} D_1 ( \tau_{42} , \tau_{32} ) + \sum_{k = 1 }^{c-1} k \tau_{41}^{ a} \tau_{42}^{ b} \tau_{31}^{ c-2} \tau_{42}^{ d} D_1 ( \tau_{31} , \tau_{31} )   +  \\  \sum_{k = 1 }^c d \tau_{41}^{ a} \tau_{42}^{ b} \tau_{31}^{ c-1} \tau_{32}^{ d-1} D_1 ( \tau_{31} , \tau_{32} )  +   \sum_{k = 1 }^{d-1} k \tau_{41}^{ a } \tau_{42}^{ b} \tau_{31}^{ c } \tau_{32}^{ d - 2 } D_1 ( \tau_{32} , \tau_{32} ).
\end{align*}
These sums can be easily done. We then get  the order $ h $ contribution to the action of the deformed Lorentz plus dilations group:
\begin{align*}
&\frac{a ( a-1)}{2} D_1 ( \tau_{41}  , \tau_{41}  ) \tau_{41}^{  a-2} \tau_{42}^{  b} \tau_{31}^{  c} \tau_{32}^{  d} +     a b D_1 ( \tau_{41}  , \tau_{42}  ) \tau_{41}^{  a-1} \tau_{42}^{  b-1} \tau_{31}^{  c} \tau_{32}^{  d} +\\ &  \frac{b ( b-1)}{2} D_1 ( \tau_{42}  , \tau_{42}  ) \tau_{41}^{  a} \tau_{42}^{  b-2} \tau_{31}^{  c} \tau_{32}^{  d}  +    b c  D_1 ( \tau_{42}  , \tau_{31}  ) \tau_{41}^{  a} \tau_{42}^{  b-1} \tau_{31}^{  c-1} \tau_{32}^{  d} +\\  &   \frac{c ( c-1)}{2} D_1 ( \tau_{31}  , \tau_{31}  ) \tau_{41}^{  a} \tau_{42}^{  b} \tau_{31}^{  c-2} \tau_{32}^{  d} +    c d  D_1 ( \tau_{31}  , \tau_{32}  ) \tau_{41}^{  a} \tau_{42}^{  b} \tau_{31}^{  c-1} \tau_{32}^{  d-1} + \\ &  \frac{d ( d-1)}{2} D_1 ( \tau_{32}  , \tau_{32}  ) \tau_{41}^{  a} \tau_{42}^{  b} \tau_{31}^{  c} \tau_{32}^{  d-2}  +    a c D_1 ( \tau_{41}  , \tau_{31}  ) \tau_{41}^{  a-1} \tau_{42}^{  b} \tau_{31}^{  c-1} \tau_{32}^{  d} +\\ &    a d  D_1 ( \tau_{41}  , \tau_{32}  ) \tau_{41}^
 {  a-1} \tau_{42}^{  b} \tau_{31}^{  c} \tau_{32}^{  d-1}  +    b d  D_1 ( \tau_{42}  , \tau_{32}  ) \tau_{41}^{  a} \tau_{42}^{  b-1} \tau_{31}^{  c} \tau_{32}^{  d-1}.
\end{align*}
This is reproduced by  the differential operator
\begin{align*}
&  \frac{ 1}{2} D_1 ( \tau_{41}  , \tau_{41}  ) \partial_{\tau_{41} }^2   +      D_1 ( \tau_{41}  , \tau_{42}  )  \partial_{\tau_{41} } \partial_{\tau_{42} }  +    \frac{1}{2} D_1 ( \tau_{42}  , \tau_{42}  )  \partial_{\tau_{42} }^2  + \\&   D_1 ( \tau_{42}  , \tau_{31}  ) \partial_{\tau_{42} } \partial_{\tau_{31} }  +    \frac{1}{2} D_1 ( \tau_{31}  , \tau_{31}  ) \partial_{\tau_{31} }^2  +      D_1 ( \tau_{31}  , \tau_{32}  ) \partial_{\tau_{31} } \partial_{\tau_{32} } + \\ &  \frac{1}{2} D_1 ( \tau_{32}  , \tau_{32}  ) \partial_{\tau_{32} }^2  +     D_1 ( \tau_{41}  , \tau_{31}  ) \partial_{\tau_{41} } \partial_{\tau_{31} }  +     D_1 ( \tau_{41}  , \tau_{32}  ) \partial_{\tau_{41} } \partial_{\tau_{32} } + \\  &    D_1 ( \tau_{42}  , \tau_{32}  ) \partial_{\tau_{42} } \partial_{\tau_{32} }.
\end{align*}
Notice that the coefficients have to match in order to get a differential operator, so the result is again non trivial. For completeness, we write the values of $D_1( \tau_{ij}  , \tau_{kl}  )$ in terms of the original variables $x, y, t$:

\begin{align*}
D_1( \tau_{41}  , \tau_{41}  )  =& -2  (y_{44}y_{43}s_{11}^{2} t_{41}t_{31}+ y_{43}^{2}s_{11}s_{21} t_{31}t_{32}+ y_{44}^{2}s_{11}s_{21}t_{41}t_{42}+\\
&  y_{44}y_{43}s_{21}^{2}t_{42}t_{32}+2y_{44} y_{43} s_{11}s_{21}t_{42}t_{31}+y_{44}y_{43}s_{11}s_{21}t_{41}t_{32}), \\
D_1( \tau_{41}  , \tau_{42}  )  =&  - (y_{43}^{2}s_{21}s_{12} t_{31}t_{32}+ y_{44}^{2}s_{21}s_{12} t_{41}t_{42}+2 y_{44}y_{43}s_{21}^{2}t_{42}t_{32+}\\
& 2 y_{44}y_{43}s_{11}s_{12}t_{41}t_{31}+2y_{44} y_{43} s_{21}s_{12}t_{42}t_{31}+y_{44}y_{43}s_{21}s_{12}t_{41}t_{32})+\\
&  y_{44}y_{43}s_{11}s_{21}t_{42}t_{31}), \\
D_1 ( \tau_{42}  , \tau_{42}  ) = &- (y_{44}^{2}s_{21}s_{12} t_{41}t_{42}+ 2y_{44}y_{43}s_{12}^{2}t_{41}t_{31}+2 y_{43}^{2}s_{12}s_{22}t_{31}t_{32}+\\
&y_{44}y_{43}s_{21}s_{12}t_{42}t_{31}+2y_{44}y_{43}s_{12}s_{22}t_{42}t_{31}+2y_{44}y_{43}
s_{12}s_{22}t_{41}t_{32}) +\\
& 3y_{44}y_{43}s_{21}s_{22}t_{42}t_{32}),\\
D_1 ( \tau_{42}  , \tau_{31}  )  =&  -  \big( y_{43} y_{33} s_{11} s_{12} t_{31}^2 + y_{44} y_{33} s_{11} s_{12} t_{41} t_{31} + 2 y_{43} y_{34} s_{11} s_{12} t_{41} t_{31} +\\ &   y_{44} y_{34} s_{11} s_{12} t_{41}^2 + y_{43} y_{34} s_{21} s_{12} t_{41} t_{32} + y_{43} y_{34} s_{21} s_{12} t_{41} t_{32} +\\  &  y_{44} y_{33} s_{11} s_{21} t_{42} t_{31} + 2 y_{44} y_{34} s_{11} s_{21} t_{41} t_{42} + y_{43} y_{33} s_{21} s_{12} t_{31} t_{32}+ \\  &   y_{43} y_{33} s_{11} s_{22} t_{31} t_{32} -2 y_{43} y_{34} s_{21} s_{12} t_{42} t_{31} + y_{43} y_{34} s_{11} s_{22} t_{42}t_{31}  +\\  &  2 y_{43} y_{34} s_{21} s_{12} t_{42} t_{31} + y_{43} y_{34} s_{11} s_{22} t_{42} t_{31} + y_{43} y_{33} s_{21} s_{22} t_{32}^2+ \\  &  2 y_{43} y_{34} s_{21} s_{22} t_{42} t_{32} \big),\\
D_1 ( \tau_{31}  , \tau_{31}  ) =&  -2  (y_{34}y_{33}s_{11}^{2} t_{41}t_{31}+ y_{33}^{2}s_{11}s_{21} t_{31}t_{32}+ y_{34}^{2}s_{11}s_{21}t_{41}t_{42}+\\
&  y_{34}y_{33}s_{21}^{2}t_{42}t_{32}+2y_{34} y_{33} s_{11}s_{21}t_{42}t_{31}+y_{34}y_{33}s_{11}s_{21}t_{41}t_{32}), \\
D_1 ( \tau_{32}  , \tau_{32}  ) =& -2  (y_{34}y_{33}s_{12}^{2} t_{41}t_{31}+ y_{33}^{2}s_{12}s_{22} t_{31}t_{32}+ y_{34}^{2}s_{12}s_{22}t_{41}t_{42}+\\
& + y_{34}y_{33}s_{22}^{2}t_{42}t_{32}+2y_{34} y_{33} s_{12}s_{22}t_{42}t_{31}+y_{34}y_{33}s_{12}s_{22}t_{41}t_{32}), \\
D_1 ( \tau_{41}  , \tau_{31}  )  =&  -  \big( y_{43} y_{34} s_{11}^2 t_{41} t_{31} + y_{43} y_{34} s_{21}^ 2 t_{42} t_{32} + 2 y_{43} y_{33} s_{11} s_{21} t_{31} t_{32} +\\ &   y_{44} y_{33} s_{11} s_{21} t_{42} t_{31} + 2 y_{43} y_{34} s_{11} s_{21} t_{42} t_{31} + y_{43} y_{34} s_{11} s_{21} t_{41} t_{32} +\\ &  2 y_{44} y_{34} s_{11} s_{21} t_{41} t_{42} \big) \\
D_1 ( \tau_{41}  , \tau_{32}  )  = &-   \big( y_{43} y_{34} s_{11} s_{12} t_{41} t_{31} + y_{43} y_{33} s_{21} s_{12} t_{31} t_{32} + y_{43} y_{34} s_{21} s_{12} t_{42} t_{31}+ \\ &   y_{43} y_{34} s_{21} s_{12} t_{42} t_{31} + y_{44} y_{34} s_{21} s_{12} t_{41} t_{42} + y_{43} y_{34} s_{21} s_{22} t_{42} t_{32}  \big), \end{align*}
\begin{align*}
D_1 ( \tau_{42}  , \tau_{32}  )  = &  -  \big( y_{43} y_{34} s_{12}^2 t_{41} t_{31} + y_{43} y_{34} s_{22}^2 t_{42} t_{32} + y_{44} y_{34} s_{21} s_{12} t_{41} t_{42} +\\ &   2 y_{43} y_{33} s_{12} s_{22} t_{31} t_{32} + 2 y_{43} y_{34} s_{12} s_{22} t_{42} t_{31} + y_{43} y_{34} s_{12} s_{22} t_{41} t_{32} \big), \\
D_1 ( \tau_{31}  , \tau_{32}  ) =&  -  \big( y_{33}^2 s_{21} s_{12} t_{31} t_{32}  + y_{34}^2 s_{21} s_{12} t_{41} t_{42} + 2 y_{34} y_{33} s_{11} s_{12} t_{41} t_{31} +\\ &  2 y_{34} y_{33} s_{21} s_{12} t_{42} t_{31} + y_{34} y_{33} s_{21} s_{12} t_{41} t_{32} +\\& y_{34} y_{33} s_{11} s_{22} t_{42} t_{31} + 2 y_{34} y_{33} s_{21} s_{22} t_{42} t_{32} \big).
\end{align*}

\section{The real forms: the Euclidean and Minkowskian signatures}\label{sec:The real}
\subsection{The real forms in the classical case}\label{subsec:therealforms}

\begin{definition}Let $\cA$ be a commutative algebra over $\C$. An involution $\iota$ of $\cA$ is an antilinear map satisfying, for $f,g\in \cA$ and $\alpha,\beta \in\C$
\begin{align}
&\iota(\alpha f+\beta g)=\alpha^*\iota f +\beta^* \iota g,\qquad &\hbox{(antilinearity)}\label{antilinearity}\\
&\iota(fg)=\iota (f)\iota(g),\qquad &\hbox{(automorphism)}\label{automorphism}\\
&\iota\circ\iota =\id.\qquad &\label{square1}
\end{align}
\hfill$\blacksquare$
\end{definition}

Let us consider the set of fixed points of $\iota$,
$$\cA^\iota =\{f\in \cA\;/\; \iota(f)=f\}.$$
It is easy to see that this is a real algebra whose complexification is $\cA$. $\cA^\iota $ is a {\it real form} of $\cA$.

\begin{example}{\sl The real Minkowski space.}

We consider the algebra of the complex Minkowski space $\cO(\M)\approx[t_{31},t_{32},t_{41},t_{42}]$ and the following involution,
$$\begin{pmatrix}\iota_\rM(t_{31})&\iota_\rM(t_{32})\\\iota_\rM(t_{41})&\iota_\rM(t_{42})\end{pmatrix}=
\begin{pmatrix}t_{31}&t_{41}\\t_{32}&t_{42}\end{pmatrix},$$ which can be also written simply as
$$\iota_\rM(t)=t^T.$$
Using the Pauli matrices (\ref{Pauli})
$$
t=\begin{pmatrix}t_{31}&t_{32}\\t_{41}&t_{42}\end{pmatrix}=x^\mu\sigma_\mu=
\begin{pmatrix}
x^{0}+x^{3} & x^{1}-\ri x^{2} \\
x^{1}+ix^{2} & x^{0}-x^{3}
\end{pmatrix},
$$
so
\begin{align*}&x^{0}=\frac 12(t_{31}+t_{42}),\qquad &x^{1}=\frac 12(t_{32}+t_{41}),\\&x^{2}=\frac 1 {2\ri} (t_{41}-t_{32}),&x^{3}=\frac 12(t_{31}-t_{42}),\\\end{align*}
are fixed points of the involution. In fact, it is easy to see that
$$\cO(\M)^{\iota_\M}=\R[x^0, x^1, x^2, x^3].$$
\hfill$\blacksquare$\end{example}

\begin{example}{\sl The Euclidean space.} We consider now the following involution on $O(\M)$
$$\begin{pmatrix}\iota_\rE(t_{31})&\iota_\rE(t_{32})\\\iota_\rE(t_{41})&\iota_\rE(t_{42})\end{pmatrix}=
\begin{pmatrix}t_{42}&-t_{41}\\-t_{32}&t_{31}\end{pmatrix}.$$ Another way of expressing it is in terms of the matrix of cofactors,
$$\iota_\rE(t)=\rcof(t).$$

The combinations
\begin{align*}&z^{0}=\frac 12(t_{31}+t_{42}),\qquad &z^{1}=\frac \ri 2(t_{32}+t_{41}),\\&z^{2}=\frac 1 {2} (t_{41}-t_{32}),&z^{3}=\frac \ri 2(t_{31}-t_{42}),\\\end{align*}
are
fixed points of $\iota_\rE$, and as before,
$$\cO(\rM)^{\iota_E}=\R[z^0, z^1, z^2, z^3 ].$$
\hfill$\blacksquare$\end{example}

We are interested now in the real forms of the complex Poincar\'{e} plus dilations that have a coaction on the real algebras. So we start with  (\ref{opl})

$$\cO(P)=\C[x_{ij},  y_{a b},  T_{ai}]/( \det x\cdot \det y -1).$$ We then look for the appropriate involution in $\cO(P)$, denoted as $\iota_{P,\rM}$ or $\iota_{P,\rE}$ `preserving' the corresponding real form of the complex Minkowski space. This means that the involution has to satisfy
\begin{align*}&\tilde \Delta\circ \iota_\rM=\iota_{P,\rM}\otimes \iota_{\rM}\circ\tilde \Delta,\\
&\tilde \Delta\circ \iota_\rE=\iota_{P,\rE}\otimes \iota_{\rE}\circ\tilde \Delta.
\end{align*}
It is a matter of calculation to check that
\begin{align}
&\iota_{P,\rM}(x)=S(y)^T, \qquad &\iota_{P,\rM}(y)=S(x)^T,\qquad &\iota_{P,\rM}(T)=T^T;\label{involutionM}\\
&\iota_{P,\rE}(x)=S(x)^T, \qquad &\iota_{P,\rE}(y)=S(y)^T,\qquad &\iota_{P,\rE}(T)=\rcof(T),\label{involutionE}
\end{align}
are the correct expressions.
It is not difficult to realize that in the Minkowskian case the real form of the Lorentz group (corresponding to the generators $x$ and $y$) is $\rSL(2,\C)_\R$ and in the Euclidean case is $\rSU(2)\times \rSU(2)$.  One can further check the compatibility of these involutions with the coproduct and the antipode
\begin{align}
&\Delta\circ\iota_{P,\rM}=\iota_{P,\rM}\otimes \iota_{P,\rM}\circ \Delta,\qquad &S\circ\iota_{P,\rM}=\iota_{P,\rM}\circ S;\label{coprodinvM}\\
&\Delta\circ\iota_{P,\rE}=\iota_{P,\rE}\otimes \iota_{P,\rE}\circ \Delta,\qquad &S\circ\iota_{P,\rE}=\iota_{P,\rE}\circ S.\label{coprodinvE}
\end{align}

\subsection{The real forms in the quantum case}

We have to reconsider the meaning of `real form' in the case of quantum algebras.
We can try to extend the involutions (\ref{involutionM}, \ref{involutionE})
to the quantum algebras. We will denote this extension with the same name since they cannot be confused in
the present context.

The first thing that we notice is that property (\ref{automorphism}) has to be modified. In fact,
the property that the involutions $\iota_{\rM}$, $\iota_{\rE}$
satisfy with respect to the commutation relations (\ref{Relations QM}) of the complex algebra $\cO(\rM)$ is that they are
antiautomorphisms, that is

\begin{align*}&\iota_{\rM}(fg)=\iota_{\rM}(g)\iota_{\rM}(f),\\
&\iota_{\rE}(fg)=\iota_{\rE}(g)\iota_{\rE}(f).
\end{align*}
This discards the interpretation of the real form of the non commutative algebra as the set of
fixed points of the involution. The other two properties are still satisfied.

When considering the involutions $\iota_{P,\rM}$ and $\iota_{P,\rE}$ in the quantum group  $\cO_q(P)$, we also obtain an antiautomorphism of algebras,
but now  the involution has to be compatible also with the
Hopf algebra structure. The coproduct is formally the same and properties (\ref{coprodinvM}, \ref{coprodinvE}) are still satisfied
(so the involutions are automorphisms of coalgebras).
On the other hand, differently from the classical case, the involutions do not commute with the antipode. This is essentially due to the fact
that $S^2\neq 1$. One can explicitly check that
\begin{align}&S^2\circ\iota_{P\rM}\circ S=S\circ\iota_{P\rM},\qquad \nonumber\\
&S^2\circ\iota_{P\rE}\circ S=S\circ\iota_{P\rE}.\label{commi}
\end{align}Property (\ref{square1}) is still satisfied, ${\iota_{P,\rM}}^2=1$ and ${\iota_{P,\rE}}^2=1$. Using this fact, (\ref{commi}) can be written as
\begin{align*}&(\iota_{P\rM}\circ S)^2=\id,\qquad \\
&(\iota_{P,\rE}\circ S)^2=\id.
\end{align*}

All these properties define what is known as a Hopf $*$-algebra structure (see for example \cite{ka}).

\begin{definition}{\sl Hopf $*$ algebra structure.} Let $\cA$ be a Hopf algebra. We say that it is a Hopf $*$-algebra
if there exists an antilinear involution $\iota$ on $\cA$ which is an antiautomorphism of algebras and an automorphism of coalgebras and such that
$$(\iota\circ S)^2=\id,$$
being $S$ the antipode.  \hfill$\blacksquare$\end{definition}
For example,  each real form of a complex Lie algebra corresponds to a $*$-algebra structure in the corresponding enveloping algebra, seen as a Hopf algebra.

\begin{remark} {Real forms on the star product algebra}
The involutions can be pulled back to the star product algebra using the quantization maps $Q_M:\cO_q(\rM)\rightarrow \cO(rM)[q, q^{-1}]$ (see (\ref{QM}), and $Q_G$ (see \ref{QG}) and then extended to the algebra of smooth functions. The Poisson bracket in terms of the Minkowski space variables ($x^\mu$) or the Euclidean ones ($z^\mu$) is purely imaginary (see \ref{pb2}), as a consequence of the antiautomorphism property of the involutions.

In the case of the quantum groups, the whole Hopf $*$-algebra structure is pulled back to the polynomial algebra and then extended to the smooth functions.
 \hfill$\blacksquare$\end{remark}

\section{The deformed quadratic invariant.}\label{sec:The deformed}

Let us consider the quantum determinant in $\cO_q(\rM)$

$$
\hat C_q= {\det}_{q}
 \begin{pmatrix}
 \hv_{32}& \hv_{31} \\
  \hv_{42} & \hv_{41}
 \end{pmatrix}
 =\hv_{32}\hv_{41}-q^{-1}\hv_{31}\hv_{42}.
 $$
Under the coaction of $\cO_q(P)$ with the translations put to zero (that is for the quantum Lorentz times dilation group), the quantum determinant satisfies

$$
\tilde{\Delta}(\hat C_{q})={{\det}_{q}\hy}\,S({{\det}_{q}\hat{x}})\otimes \hat C_{q},
$$
so if we suppress the dilations, then ${\det}_{q}\hy=1$, ${\det}_{q}\hat{x}=1$ and the determinant is a {\it quantum invariant},
$$
\tilde{\Delta}(\hat C_{q})=1\otimes \hat C_{q}.
$$

The invariant $\hat C_q$ can be pulled back to the star product algebra with the quantization map $Q_\M$:
\be C_q=Q_\M^{-1}(\hat C_q )=t_{41}t_{32}-qt_{42}t_{31}.\label{spm1}\ee

We can now change to the Minkowski space variables, and the quadratic invariant in the star product algebra is
\be C_q=-q (x^0)^2+q(x^3)^2+(x^1)^2+(x^2)^2.\label{spm2}\ee
$C_q$ is the quantum star invariant. Notice that the expressions (\ref{spm1},\ref{spm2}) depend upon the quantization map or ordering rule chosen.

\section{Conclusions.} \label{sec:Conclusions}

In this paper we have computed an explicit formula for a star product on polynomials on the complexified  Minkowski space. This star product has several properties:

\begin{itemize}

\item It can be extended to a star product on the conformal space $G(2,4)$. This is done by gluing the star products computed in each open set (\ref{opensets}).

\item It can be extended to act on smooth functions as a differential star product.

\item The Poisson bracket is quadratic in the coordinates.

\item There is a coaction of the quantum Poincar\'{e} group (or the conformal group in the case of the conformal spacetime) on the star product algebra.

\item   It has at least two real forms corresponding to the Euclidean and Minkowski signatures.

\item It can be extended to the superspace (to chiral and real superfields).

\end{itemize}

Since fields are smooth functions, the differentiability of the star product gives a hope that one can develop a quantum deformed field theory, that is, a field theory on the quantum deformed Minkowski space. The departure point will be to find a generalization of the Lapacian and the Dirac operator associated to the quantum invariant $C_q$.

One advantage of using the quantum group $\rSL_q(4,\C)$ is that the coalgebra structure is isomorphic to the coalgebra of the classical group $\rSL(4,\C)$ (see for example Theorem 6.1.8 in Ref. \cite{cp}). This means that the group law is unchanged, so the Poincar\'{e} symmetry principle of the field theory would be preserved in the quantum deformed case.

\section*{Acknowledgments}

D. Cervantes wants to thank the Departament de F\'{\i}sica Te\`{o}rica,
Universitat de Val\`{e}ncia for the hospitality
during the elaboration of this work.

Felip A. Nadal wants to thank CSIC for a JAE-predoc grant.

This work has been supported in part by grants FIS2008-06078-C03-02 of Ministerio de Ciencia e Innovaci\'{o}n (Spain), FIS2011-29813-C02-02, FIS2014-57387-C3-1 and SEV-2014-0398 of Ministerio de Econom\'{\i}a y Competitividad and  ACOMP/2010/213 and ACOMP/2013/179 of Generalitat Valenciana.

\appendix

\section{A basis for the Poincar\'{e} quantum group}\label{diamond}

In this appendix we  prove that, given a certain specific ordering on the
generators of the Poincar\'{e} quantum group, the ordered monomials
form a basis for its quantum algebra. This is a non trivial result
based on the classical work by G. Bergman \cite{be}.

\subsection{Generators and relations for the Poincar\'{e}
quantum group}
\label{genrelpoincare}

Let us consider $\rSL_q(n,\C)$ the quantum complex general linear group
with indeterminates $g_{IJ}$ subject to the Manin relations (\ref{ManinR}) and (\ref{qdet}\footnote{In this appendix we write the noncommutative generators without the hat `$\hat {}$' to simplify the notation.}.
(see \cite{cfl2} sec. 7)\footnote{All of the arguments
in this appendix hold replacing $\rSL_q(n,\C)$ with the general
linear quantum group and the complex field with any field of
characteristic zero.}.
Inside $\rSL_q(n,\C)$ we consider the following elements, which we
write, as usual, in a matrix form:

\begin{align*}
&x =\begin{pmatrix}
g_{11} & g_{12}\\ g_{21} & g_{22}  \end{pmatrix},
&&
T =
\begin{pmatrix} -q^{-1}D_{23}D_{12}^{-1} & D_{13}D_{12}^{-1}\\
-q^{-1}D_{24}D_{12}^{-1} & D_{14}D_{12}^{-1}  \end{pmatrix}
\nonumber \\\nonumber \\
&y = \begin{pmatrix}
g_{33} & g_{34}\\ g_{43} & g_{44} \end{pmatrix}.&&
\\ \\
\end{align*}

As in (\ref{qpoincare}), let us define the \textit{quantum Poincar\'{e} group}, $\cO_q(P)$ as the subring
of  $\rSL_q(n,\C)$ generated by the elements in the matrices $x$, $y$, $T$
defined above.
In order to give a presentation for $\cO_q(P)$ we need to consider
all of the commutation relations between the generators (\ref{commx}, \ref{commy}, \ref{commT}, \ref{commxt}, \ref{commyt}).

The entries in $x$ (resp. $y$) satisfy  the Manin
commutation relations in dimension 2, that is,
\begin{align*}&x =\begin{pmatrix}
g_{11} & g_{12}\\ g_{21} & g_{22}  \end{pmatrix}\sim\begin{pmatrix}
a & b\\ c & d  \end{pmatrix},\qquad y = \begin{pmatrix}
g_{33} & g_{34}\\ g_{43} & g_{44} \end{pmatrix}\sim  \begin{pmatrix}
a & b\\ c & d  \end{pmatrix}\\\\
&ba=qab,\qquad  ca=qac, \qquad
db=qbd, \qquad  dc=qcd, \\ &cb=bc \qquad  da=ad-(q^{-1}-q)bc.
\end{align*}
Moreover, they commute with each other:
$$
x_{IJ}y_{KL}=y_{KL}x_{IJ}.
$$
Similarly one can show that the entries in $T_{IJ}$ satisfy the
Manin relations, with the order
 $$T =\begin{pmatrix}
T_{32} & T_{31}\\ T_{42} & T_{41}  \end{pmatrix}\sim  \begin{pmatrix}
a & b\\ c & d  \end{pmatrix},$$ but they do not commute with $x$ and $y$ (\ref{commxt}, \ref{commyt}).

This provides a presentation of $\cO_q(P)$ in terms of
generators and relations (\ref{presentation oqpl}) (see \cite{cfl2} for more details),
$$\cO_q(P)=\C_q\langle  x_{IJ},  y_{KL},   T_{RS}\rangle/(\cI_{P}, \; {\det}_q  x \cdot {\det}_q y-1),$$
where $\cI_{P}$ is the ideal generated by the commutation relations (\ref{commx}, \ref{commy}, \ref{commT}, \ref{commxt}, \ref{commyt}).

\subsection{The Diamond Lemma}

Let us recall some definitions and theorems from the
fundamental work by Bergman \cite{be}
(see also \cite{ks} pg 103)
\footnote{All of our arguments hold more in general
replacing $\C_q$ with a commutative ring with $1$.}.

\begin{definition}
Let $\C_q\langle x_i\rangle$ be the free associative algebra over $\C_q$
with generators $x_1, \dots , x_n$
and let
$$
X:=\{X_I=x_{i_1} \cdots x_{i_s} \; / \; I=(i_1, \dots ,i_s), \;
i_j \in \{1, \dots, n\}\}
$$
be the set of all (unordered) monomials. $X$ is clearly a basis
for $\C_q \langle x_i\rangle$. We define on $X$ an order, $<$, such that given two monomials
$x$ and $y$, then $x < y$ if the length of $x$ is less  than the length of $y$
and for equal lengths  we apply the lexicographical ordering. \hfill $\blacksquare$

\end{definition}

Let $\Pi=\{(X_{I_k}, f_k) \; | \; k=1, \dots, s\}$ be a certain set of pairs
$X_{I_k} \in X$ and $f_k \in  \C_q\langle x_i\rangle$. We denote by $\cJ_\Pi$ the ideal
$$
\cJ_\Pi=(X_{I_k}-f_k, \; k=1, \dots, s) \subset \cO_q(P).
$$
In our application $\Pi$ will
yield the ideal  of the commutation relations for the
quantum Poincar\'{e} group.

\begin{definition} We say that $\Pi$ is \textit{compatible} with
the ordering $<$ if $f_k$ consists of a linear combination
of ordered monomials.\hfill $\blacksquare$
\end{definition}

For example if $M_q(2)=\C_q\langle a,b,c,d\rangle/\cI_M$, where $\cI_M$ is the ideal
of the Manin relations, we have that
$$
\Pi_M=\{(ba, qab), \, (ca, qac), \,
(cb, bc), \, (dc, qcd), \, (db, qbd), \, (da, ad-(q^{-1}-q)bc) \,\}
$$
is compatible with the ordering $a<b<c<d$.

We want to find a basis consisting of ordered
monomials for a $\C_q$-module $\C_q \langle x_i\rangle/\cJ_\Pi$. Clearly
this is not  possible for any chosen total order. However, when $\Pi$ is compatible with the order, that is, when the relations $X_{I_k}-f_k$
behave nicely with respect to the given order, then  we can device an algorithm
to reduce any monomial to a \textit{standard form}
(namely to writing it as a combination of ordered monomials). This is essentially the content of the Diamond Lemma for ring theory
that we shall describe below.

We have  two problems to solve: first, one has to make sure that
any procedure to reduce a monomial to the standard form terminates,
and then one has to make sure that the chosen procedure gives a unique
result.

\begin{definition}
Assume that we fix a generic set $\Pi$ as above.
Let $x,y \in X$ and let $r_{xky}$ be the linear map of
$\C_q\langle x_i\rangle $ sending the elements of the form
$xx_{i_k}y$ to $xf_ky$ and leaving the rest unchanged. $r_{xky}$ is
called a \textit{reduction} and an element $x \in X$ (or more
generally in $\C_q\langle x_i\rangle$) is \textit{reduced} if $r(x)=x$ for all
reductions $r$. \hfill $\blacksquare$
\end{definition}

In general more than one reduction can be applied to an element.
For example if we take the quantum matrices $M_q(2)$ and $\Pi_M$ as
above, we see that $dcba$ is not reduced, and we have several ways to
proceed to reduce it. We want to make sure that that there are
no ambiguities, or, 
in other words,
we want to make sure there is a unique reduced
element associated with it.

\begin{definition}
Let $x,y,z \in X$ and $x_{i_k}$, $x_{i_l}$ be the first elements
of two pairs in $\Pi$. We say that $(x,y,z,x_{i_k}, x_{i_l})$
form an \textit{overlapping ambiguity} if  $x_{i_k}=xy$, $x_{i_l}=yz$.
The ambiguity is \textit{resolvable} if there are two reductions
$r$ and $r'$ such that $r(x_{i_k}z)= r'(xx_{i_l})$. In other words, if
we can reduce $xyz$ in two different ways, we must obtain the same result.
Similarly $(x,y, x_{i_k}, x_{i_l})$ form an \textit{inclusion ambiguity}
if $x_{i_k}=xx_{i_l}y$. The inclusion ambiguity is \textit{solvable}
if  there are two reductions
$r$ and $r'$ such that $r(x_{i_k})= r'(xx_{i_l}y)$.
 \hfill $\blacksquare$
\end{definition}

\begin{theorem} \label{diamondlemmathm} (Diamond Lemma).
Let $R$ be the ring defined by generators and relations as:
$$
R:=\C_q\langle x_i\rangle /(X_{I_k}-f_k,k=1 \dots s)
$$
If $\Pi=\{X_{I_k},f_k\}_{k=1, \dots, s}$ is compatible
with the ordering $<$ and all ambiguities are resolvable, then
the set of ordered monomials is a basis for $R$. Hence $R$ is
a free module over $\C_q$.
\end{theorem}

{\sl Proof}.  See \cite{be}.

\hfill $\blacksquare$

\medskip

\subsection{A basis for the Poincar\'{e} quantum group}


In this section, we want to apply the Diamond Lemma, to obtain
an explicit basis for the quantum algebra of the Poincar\'{e}
quantum group.
Let us fix a total order on the variables $x$, $y$, $t$ as follows:
$$
t_{32} \, > \, t_{31} \, > \, t_{42} \, > \, t_{41}
\, > \, x_{11} \, > \, x_{12} \, > \, x_{21} \, > \, x_{22} \, > \,
y_{33} \, > \, y_{34} \, > \, y_{43} \, > \, y_{44}.
$$

One sees right away that the relations in $\cI_M$ as described in
(\ref{commx}, \ref{commy}, \ref{commT}, \ref{commxt}, \ref{commyt}) give raise to a $\Pi$ compatible with the given
order. Furthermore, notice that this order is the Manin ordering
 (see \cite{ma}) in two dimensions when restricted
to each of the sets $\{x_{IJ}\}$, $\{y_{KL}\}$, $\{t_{RS}\}$.

\medskip

As one can readily see, the fact that $\Pi$ is compatible
with the given order ensures that any reordering procedure terminates.

\begin{theorem} \label{diamlemmapoincare}
Let  $\cO_q(P)=\C_q\langle x_{ij},y_{kl}, t_{il}\rangle /\cI_{P}$
be the algebra corresponding to the quantum Poincar\'{e}
group. Then, the monomials
in the order:
$$
t_{32} \, > \, t_{31} \, > \, t_{42} \, > \, t_{41}
\, > \, x_{11} \, > \, x_{12} \, > \, x_{21} \, > \, x_{22} \, > \,
y_{33} \, > \, y_{34} \, > \, y_{43} \, > \, y_{44}.
$$
are a basis for $\cO_q(P)$.
\end{theorem}

{\sl Proof}. By the Diamond Lemma \ref{diamondlemmathm} we only need
to show that all ambiguities are resolvable.
We notice that when two generators $a$, $b$, $q$-commute, that is
$ab=q^sba$, they behave, as far the reordering is concerned,
exactly as commutative indeterminates.
Hence we only take into consideration ambiguities where no
$q$-commuting relations appear. The proof consists in checking
directly that all such ambiguities are resolvable.

Let us see, as an example of the procedure to follow,
how to show that the ambiguity $x_{22}x_{11}t_{32}$ is
resolvable. All the other cases follow the same pattern
since the relations have essentially the same form as far
as the reordering procedure is concerned.

We shall indicate the application of a reduction with an arrow, as
it is customary to do.

$$
\begin{array}{rl}
(x_{22}x_{11})t_{32} \, & \lra \, (x_{11}x_{22}-(q^{-1}-q)x_{12}x_{21})t_{32} \,
\lra x_{11} (q^{-1}t_{32}x_{22}+ \\ \\
& + (q^{-1}-q)t_{31}x_{12})-
(q^{-1}-q)[x_{12}(q^{-1}t_{32}x_{21}+(q^{-1}-q)t_{31}x_{11})] \\ \\
& \lra q^{-1}t_{32}x_{11}x_{22}+q^{-1}(q^{-1}-q)t_{31}x_{11}x_{12}+ \\ \\
&-q^{-1}(q^{-1}-q)t_{32}x_{12}x_{21}-q(q^{-1}-q)t_{31}x_{11}x_{12}= \\ \\
&=q^{-1}t_{32}x_{11}x_{22}-q^{-1}(q^{-1}-q)t_{32}x_{12}x_{21}+
(1-q^2)t_{31}x_{11}x_{12}.
\end{array}
$$
Similarly
$$
\begin{array}{rl}
x_{22}(x_{11}t_{32}) \, & \lra \, x_{22}t_{32}x_{11} \, \lra \,
(q^{-1}t_{32}x_{22}+(q^{-1}-q)t_{32}x_{12})x_{11} \\ \\
& \lra \, q^{-1}t_{32}(x_{11}x_{22}-(q^{-1}-q)x_{12}x_{21})+
(1-q^2)t_{31}x_{11}x_{12}.
\end{array}
$$

As one can see the two expressions are the same and reduced, hence we obtain
that this ambiguity is resolvable.
\hfill $\blacksquare$

\medskip

\begin{remark}
We end the discussion by noticing that the Theorem
\ref{diamlemmapoincare} holds also for
the order:
$$
x_{11} \, > \, x_{12} \, > \, x_{21} \, > \, x_{22} \, > \,
y_{33} \, > \, y_{34} \, > \, y_{43} \, > \, y_{44} \,>\,
t_{32} \, > \, t_{31} \, > \, t_{42} \, > \, t_{41}
$$
the proof being the same.\hfill $\blacksquare$
\end{remark}

\end{document}